\documentclass[full paper,twocolumn]{jpsj2}

\title{%
Geometrical aspects of Gigantic Magneto-Electric effect and Quantum
Pump }
\author{{Ryuichi Shindou}$^{1}$\thanks{E-mail address: 
shindou@appi.t.u-tokyo.ac.jp.}, 
{Naoto Nagaosa}$^{1,2}$
\thanks{E-mail adress: nagaosa@appi.t.u-tokyo.ac.jp}}

\inst{$^{1}$ Department of Applied Physics, University of Tokyo,
Bunkyo-ku, Tokyo 113-8656, Japan\\
$^{2}$ CERC, AIST Tsukuba Central 4, Tsukuba 305-8562, Japan}

\recdate{\today}

\abst{%
We study the Magneto-Electric 
(ME) effect from the viewpoint of the Berry phase connection 
and quantum adiabatic charge transport . 
The linear response theory for the electronic polarization 
$\vec{P}_{\rm{el}}$ can be interpreted in terms of the flux 
or the fictitious magnetic field related to the Berry phase in 
the generalized momentum space including external parameters.
An applied magnetic field modifies the spin 
configuration, induces adiabatic deformation of the Bloch 
wavefunction, and results in the electronic polarization.   
This provides a new mechanism for the gigantic ME effect.
For a cyclic change of the applied magnetic field, even a quantized 
charge transport is possible.}

\kword{%
Magneto Electric effect, macroscopic electronic polarization, 
spin ordering field, generalized momentum space, 
fictitious magnetic field, magnetic monopole, Berry phase, 
depolarization field}
\begin{document}
\maketitle
\section{Introduction}
The interplay between spin and charge degrees of freedom
is one of the central issues in the physics of strongly correlated
electronic systems. From this viewpoint, the cross correlation between
these two degrees of freedom is of particular interests.
The Magneto-Electric (ME) effect is a phenomenon where 
a finite magnetization $\vec M$ appears under an applied 
electric field $\vec E$ or an electric 
polarization $\vec P$ is induced by an 
external magnetic field $\vec H$.

This effect has been studied for a long term since its
theoretical prediction\cite{dzyaloshinskii}
and experimental observations\cite{rado} in ${\rm{Cr_{2}O_{3}}}$.
A phenomenological description of this effect is given by the
free energy containing $\alpha_{ij}E_{i}H_{j}$.
The selection rule for the ME coefficient $\alpha_{ij}$ is
obtained from the group theoretical considerations.
In the case of antiferromagnetically
ordered ${\rm{Cr_{2}O_{3}}}$, the time-reversal 
symmetry ($R$) accompanied by the spatial translation
($T$) is broken due its peculiar crystal structure, 
in contrast to usual antiferromagnets.
Instead, its AF ordered phase
is invariant under the product of the time-reversal symmetry ($R$)
and the spatial inversion ($I$), which does not prohibit the 
finite matrix element $\alpha_{ij}$.
 Furthermore, due to the several other symmetries in
antiferromagnetically
ordered ${\rm{Cr_{2}O_{3}}}$, there remain only the diagonal
matrix elements $\alpha_{xx}=\alpha_{yy}=\alpha_{\perp},\alpha_{zz}
=\alpha_{\parallel}$\cite{dzyaloshinskii}.

Microscopic theories of the ME effect in $\rm{Cr_{2}O_{3}}$ based
on these phenomenological observations
have been proposed\cite{alexander,date,yatom}. According to these theories,
the applied electric field shifts the oxygen ions $\rm{O^{2-}}$ and
breaks the crystallographic equivalence between several sublattices composed
of ${\rm{Cr}}$ atoms. For example, the g-factors on two ${\rm{Cr}}$
sublattices differ from each other due to the crystal field produced by
the shifted oxygen ions. This difference induces a finite net
magnetization\cite{alexander}.
A temperature dependence of $\alpha_{\parallel}$ seems to be well
explained by this mechanism \cite{alexander,yatom}.
When we trace these theories the other way around, a microscopic origin
for the {\it{electric polarization}} induced by the applied magnetic field
was attributed to the displacement of the ligand anion atoms,
namely the lattice displacements.

Instead, we are going to study a 
{\it{Bloch electron's contribution}} to the 
macroscopic electronic polarization (MEP) induced by the applied 
magnetic field. 
The MEP in dielectrics has been formulated only recently,
since an electron's position operator is ill-defined
in the Hilbert space spanned by the Bloch wavefunctions which obey
the periodic Born-von Karman (BvK) boundary condition
\cite{vanderbilt,resta}.
In 1990s, it is gradually recognized that the MEP 
in ferroelectrics is the average value of    
the total charges transferred through the unit area 
in the system associated with the atomic displacement\cite{resta}.  
The atomic displacement and/or 
lattice distortions are induced also by 
the external pressure, leading to 
the piezo-electric effects\cite{vanderbilt1,avron}.

The MEP is closely related to the quantized adiabatic 
particle transport proposed 
by Thouless and Niu in the 1980s\cite{thouless}. 
They considered a 1D band insulator with the periodic
 boundary condition. When some external parameters 
are deformed adiabatically, and are put back to 
their initial values, the particle number transferred during this 
cycle is quantized to be non-zero integer. 
This means that the MEP increases by a finite amount, even 
though the external parameters get back to their initial 
values. Recently this kind of ``d.c.'' responses 
induced by ``a.c.'' impulses, which are generally called 
Quantum Pump, are extensively studied in 
meso (nano)-scale systems, where a cyclic variation of a
potential shape induced by the gate voltage pumps up 
electrons from one end to the other end of the system
\cite{switkes,altshuler,brouwer}.
However we should make a clear distinction between 
the Quantum Pump  discussed in these mesoscale systems 
and that of the 1D band insulator. Namely, the former 
systems is an open system attached to the leads with 
finite $S$-matrix elements between two channels, 
while, in the latter system, the wavefunction in one end 
and that of the other do not overlap due to the presence of 
the gapped region between them. Futhermore, in the latter 
system, the particle number transferred through this gapped 
region during a cyclic process becomes quantized in 
the thermodynamic limit \cite{thouless,niu,shindou}, which 
is not the case with the mesoscale systems. 

Both MEP and Quantum Pump are closely related to the Berry phase.
Berry phase is the quantal phase 
associated with an adiabatic time-evolution 
of a wavefunction. Let an external parameter 
evolve from $\vec{\lambda}_{i}$ to $\vec{\lambda}$ 
along a path $\Gamma$\cite{berry}. Then the adiabatically 
evolved wavefunction $|\Psi(\vec{\lambda}(t),t)\rangle$ 
is given as follows.   
\begin{eqnarray}
&&|\Psi(\vec{\lambda},t)\rangle=\nonumber \\
&&e^{-\int_{\Gamma}
^{\vec{\lambda}_{i} \rightarrow \vec{\lambda}(t)}
\langle \Psi_{\rm{inst}}(\vec{\lambda}^{\prime})|
\vec{\nabla}_{\vec{\lambda}^{\prime}}
|\Psi_{\rm{inst}}(\vec{\lambda}^{\prime})\rangle 
\cdot d\vec{\lambda}^{\prime}}
|\Psi_{\rm{inst}}(\vec{\lambda}(t))\rangle, \label{1-1}
\end{eqnarray} 
where $|\Psi_{\rm{inst}}(\vec{\lambda})\rangle$ is 
an instantaneous eigenstate of the corresponding 
Hamiltonian $\hat{H}(\vec{\lambda})$ ,
 and we take an initial  
wavefunction $|\Psi(\vec{\lambda}(t_{i})\equiv
\vec{\lambda}_{i},t_{i})\rangle$ 
 as $|\Psi_{\rm{inst}}(\vec{\lambda}_{i})\rangle$. 
The r.h.s. of eq.(\ref{1-1}) is invariant under the U(1) 
gauge transformation;  
$|\Psi_{\rm{inst}}(\vec{\lambda})\rangle\rightarrow
e^{i\theta(\vec{\lambda})}
|\tilde{\Psi}_{\rm{inst}}(\vec{\lambda})\rangle$ where 
$\theta(\vec{\lambda}_{i})=0$. Futhermore,  
when this parameter evolves along a closed loop 
$\Gamma_{\rm cyc}$,  a total phase factor this w.f. 
acquires, which is called U(1) phase holonomy, 
can be given by the surface-integral of the {\it{flux}} 
over an arbitrary 
surface $S$ whose boundary is the loop $\Gamma_{\rm{cyc}}$; 
\begin{eqnarray}
e^{-\oint_{\Gamma_{\rm{cyc}}}
\langle \Psi_{\rm{inst}}(\vec{\lambda})|\vec{\nabla}_{\vec{\lambda}}
|\Psi_{\rm{inst}}(\vec{\lambda})\rangle \cdot d\vec{\lambda}}
= e^{- i\int_{S}\vec{\cal B}(\vec{\lambda})\cdot d\vec{S}}, \nonumber 
\end{eqnarray} 
where the flux $\vec{\cal B}(\vec{\lambda})$  is defined as follows; 
\begin{eqnarray}
\vec{\cal B}(\vec{\lambda})\equiv -i\vec{\nabla}_{\vec{\lambda}}\times 
\langle \Psi_{\rm{inst}}(\vec{\lambda})|
\vec{\nabla}_{\vec{\lambda}}
|\Psi_{\rm{inst}}(\vec{\lambda})\rangle. \label{1-2}
\end{eqnarray}  
Since this flux is independent of the gauge choice of the 
instantaneous eigenstate, $\vec{\cal B}(\vec{\lambda})$ 
is also called {\it{fictitious magnetic field}}, 
while $\vec{\cal A}(\lambda)
\equiv\langle \Psi_{\rm{inst}}(\vec{\lambda})|
\vec{\nabla}_{\vec{\lambda}}
|\Psi_{\rm{inst}}(\vec{\lambda})\rangle$   
 corresponds to its associated vector potential. 

This kind of the flux also appears in the context of 
the MEP and/or the Quantum Pump, 
since the current is represented by the derivative 
of the phase of the wavefunctions. 
Specifically the integrated particle current can be 
viewed as the flux  penetrating a plaquette spanned by 
the crystal momentum and the external parameter. 
Although this definition of the flux  
 is slightly modified from that of eq.(\ref{1-2})   
(see eqs.(\ref{2-8},\ref{2-9})),  both of these 
are essentially the same mathematical objects, 
which are called  field strengths in the framework of  
the gauge theory. Futhermore, the point 
where the energy degeneracy occurs plays a role 
of the source of the flux, so called, 
magnetic monopole (fictitious magnetic charge). 
Generally speaking, a magnetic monopole is an origin of 
nontrivial structures of the Fiber bundle 
associated with phase factors of wavefunctions\cite{nakahara}.

Recently the spontaneous (anomalous) Hall effect observed 
in ferromagnets are discussed in the context of 
the integer quantum Hall current from the viewpoint 
of the Berry phase\cite{taguchi}. 
For example, the Hall conductivity $\sigma_{xy}$, 
is given by the total flux penetrating a plaquette spanned by 
$x$-component and $y$-component of the crystal momenta
(2D magnetic Brillouin zone)\cite{TKNN}. 
Throughout these extensive studies, it turned 
out that non-coplanar spin 
configurations and/or spin-orbit couplings 
in magnets generate magnetic monopoles and 
thus nontrivial distributions of the flux in their  
crystal momentum spaces, which result in their  
anomalous Hall currents.

In this paper, we study the MEP induced by 
a deformation of background spin configurations 
in the magnetic materials. Since the spin ordering fields 
(external parameters) in the magnetic systems can 
be controlled by the applied magnetic field, 
this corresponds to the electronic contribution to the ME effect. 
As discussed above, the macroscopic electronic polarization (MEP) 
and the transverse conductivity $\sigma_{xy}$ are closely related. 
Namely, these two physical quantities 
are the different components of 
the same field in the generalized momentum space.
Through the analogy to the integer quantum Hall effect,
we can naturally expect that even the quantized charge 
transport might be possible during the cyclic change of 
spin ordering fields. In fact, both of them are 
related to the first Chern numbers associated 
with the filled bands. 

The main results obtained in this 
paper are summarized as follows. 
From the viewpoint of the perturbation 
theory,  the Magneto-Electric (ME) effect was known to be enhanced 
when the energy denominator is small, i.e., when the band gap is 
small. However we propose, in this paper, a new mechanism 
of the ME effect, where not only the band gap reduction  
but also the U(1) phase associated with the 
magnetic Bloch wavefunction play  
very important roles in determining the magnitude of the ME effect. 
From this viewpoint, we can classify the magnetic dielectrics 
into two categories, i.e., with and without the nontrivial 
structure  of the U(1) Fiber bundle associated with magnetic 
Bloch wavefunctions. Former category is a good candidate to which
 our mechanism can be applied. In these systems, 
a nontrivial topological structure is originated from 
the band crossing located in the parameter space. 
As for a specific example of this  
new category of magnetic dielectrics, we construct a model 
where such a non-trivial topological structure 
is realized in its generalized momentum space and  
gigantic electronic polarizations are 
induced by the applied magnetic field. In fact, their 
magnitudes amount to
the order of $e/{\rm a}^{2}\sim 1 \ [{\rm{C/m^{2}}}]$, 
where {\it{e}} is the electron charge and 
 we take the lattice constant 
``${\rm{a}}$'' to be about $4{\rm \AA}$.
A typical ME dielectric ${\rm{Fe}}_{3}{\rm{O}}_{4}$ 
shows an electric polarization 
of $\sim 1.0\times10^{-5}\ [{\rm{C/m^{2}}}]$
under an applied magnetic field $H=5\ [{\rm{kOe}}]$ 
\cite{rado1}.
Compared with this, the ME effect discussed in this paper 
has 5 orders of magnitude larger (Sec. III).
Thus these electronic contribution to the ME effects 
can be hardly neglected compared 
with that of the displacement of the ligand anion 
atoms and so on, although the electronic contribution to 
the ME effects has been believed to be much smaller 
compared with the latter. 

This paper is organized as follows.
In section 2, we give our general idea 
for the Magneto-Electric effect (ME) based on the  
geometrical viewpoints and explain how the 
flux (fictitious magnetic field) is introduced 
in the context of the spin dependent MEP.  
In section 3, we propose a model which 
has a nontrivial topological structure in its generalized 
momentum space and exhibits a gigantic ME response. 
Those who want to get the overview of 
this paper can skip this section. When a system is terminated 
without electrodes, our gigantic ME effect, 
as well as conventional ME effects, suffers from depolarization 
fields. In section.4, we give an appropriate mean-field 
arguments on this problems in the context of spin dependent 
MEP. Section 5 is devoted to conclusions.                   

\section{Geometrical viewpoints of the spin dependent MEP}
\subsection{Macroscopic Electronic Polarization}
In this paper, we study the change of the electronic
polarization induced by a perturbation $\delta \hat{H}$ 
added to an original Hamiltonian $\hat{H}$: 
\begin{eqnarray}
\delta \hat{P}_{{\rm el},\mu}\equiv -\frac{e}{V}
(\hat{X}_{{\rm el},\mu}
-\langle\hat{X}_{{\rm el},\mu}\rangle_{\hat{H}}).
\end{eqnarray}
Here $\hat{X}_{{\rm el},\mu}$ is the sum of all 
electron's position operators and $V$ is the volume of the system.
$\langle\hat{X}_{{\rm el},\mu}\rangle_{\hat{H}}$ is an expectation
value of $\hat{X}_{{\rm el},\mu}$ before the
perturbation is introduced. According to the Kubo
formula for the linear response, $\delta P_{{\rm el},\mu}\equiv \langle
\delta \hat{P}_{{\rm el},\mu}\rangle_{\hat{H} + \delta \hat{H}}$
is given as 
\begin{eqnarray}
\delta P_{\rm el}
=-i\frac{e}{V}\sum_{m \ne 0}\left(
\frac{\langle \Psi_{0}|\hat{X}_{{\rm el},\mu}|\Psi_{m} 
\rangle\langle \Psi_{m}|
\delta \dot{\hat{H}} |\Psi_{0} \rangle}
{(E_{m}-E_{0})^{2}} - {\rm{c.c.}}\right)\label{2-1},
\end{eqnarray}
where $\delta \dot{\hat{H}}=i[\hat{H},\delta \hat{H}]$
and $|\Psi_{m} \rangle$ is an eigenstate of the Hamiltonian 
$\hat{H}$ with its eigenenergy $E_{m}$. 
Here we assume a finite energy gap between 
the ground state $|\Psi_{0}\rangle$  and the excited
state $|\Psi_{m} \rangle\ (m \ge 1)$ and focus on the 
zero-temperature case. When we transfer the time derivative from 
$\delta \hat{H}$ to $\hat{X}_{{\rm el},\mu}$;  
\begin{eqnarray}
&&\langle \Psi_{0}|\hat{X}_{{\rm el},\mu}
|\Psi_{m} \rangle\langle \Psi_{m}|
[\hat{H},\delta \hat{H}] |\Psi_{0} \rangle \nonumber \\
&&\ \  =\ -\langle \Psi_{0}|[\hat{H},\hat{X}_{{\rm el},\mu}]
|\Psi_{m} 
\rangle\langle \Psi_{m}
|\delta \hat{H} |\Psi_{0} \rangle,\label{2-1-0} 
\end{eqnarray}
we obtain the following expression for $\delta \vec{P}_{\rm el}$,
\begin{eqnarray}
\delta P_{{\rm el},\mu}
= -\frac{i}{V}\sum_{m\ne0}\left[
\frac{\langle\Psi_{0}|\hat{J}_{{\rm el},\mu}
|\Psi_{m}\rangle\langle\Psi_{m}|
\delta \hat{H}|\Psi_{0}\rangle}{(E_{m}-E_{0})^{2}}- {\rm{c.c.}}
\right].\label{2-4}
\end{eqnarray}
Here we introduced an electronic current operator by   
$\hat{J}_{{\rm el},\mu}=-ie[\hat{H},\hat{X}_{{\rm el},\mu}]$. 
As shown in Appendix A, this expression for the  
electronic polarization $\delta \vec{P}_{\rm{el}}$ 
can be also derived as an integrated electron current 
over a period during which the perturbation $\delta H$ 
is introduced. In the followings, we use eq.(\ref{2-4}) instead 
of eq.(\ref{2-1}). This is because the current 
operator $\hat{J}_{{\rm el},\mu}$ is well-defined on 
a periodic lattice and thus easy to deal with,  
while the position operator $\hat{X}_{{\rm el},\mu}$ 
is not.

Let the Hamiltonian depend on some external parameters
$\vec{\varphi}$, i.e. $\hat{H}(\vec{\varphi})$. Then,  
by considering $\delta \hat{H}$ in the above formalism  
as $\hat{H}(\vec{\varphi}+\delta \vec{\varphi})
-\hat{H}(\vec{\varphi})$, we obtain a derivative of 
the electronic polarization $\vec{P}_{\rm{el}}$
with respect to $\vec{\varphi}$ as
\begin{eqnarray}
\frac{\partial P_{{\rm el},\nu}}{\partial \varphi_{\mu}}
=-\frac{i}{V}\sum_{m \ne 0}\left(
\frac{\langle \Psi_{0}|\hat{J}_{{\rm el},\nu}|\Psi_{m} \rangle
\langle \Psi_{m}|
\frac{\partial \hat{H}}{\partial \varphi_{\mu}}|\Psi_{0} \rangle}
{(E_{m}-E_{0})^{2}} - {\rm{c.c.}}\right).\label{2-5}
\end{eqnarray}

\subsection{spin dependent MEP and the Flux}
Now we study the spin dependent MEP. In this case,
we regard $\vec{\varphi}$ as the spin ordering field
$\vec{\phi}_{i,{\rm SP}}$ as follows.
At first we decouple the on-site Coulomb repulsion by using 
the Stratonovich-Hubbard field $\vec{\phi}$ 
and replace it by its saddle point 
solution $\vec{\phi}_{i,{\rm SP}}$ as,
\begin{eqnarray}
&&Z[\vec{h}]\approx\int{\cal{D}}C^{\dagger}{\cal{D}}C
{\rm{exp}}\biggl[ -\int L(\vec{\phi}_{\rm SP}
,C^{\dagger},C,\vec{h}) \biggr],
\nonumber  \\
&&L(\vec{\phi}_{i,{\rm S.P.}},C^{\dagger},C,\vec{h}) = \nonumber \\
&& \ \frac{U}{4}\sum_{i}|\vec{\phi}_{i,{\rm SP}}|^{2}
+\sum_{i,\alpha}C^{\dagger}_{i,\alpha}({\partial}_{\tau}-\mu)
C_{i,\alpha} + H_{\rm M.F.} \nonumber \\
&&H_{\rm M.F.}= \nonumber \\ 
&& - \  \sum_{i,j,\alpha}t_{ij}C^{\dagger}_{i,\alpha}C_{j,\alpha}
+\frac{U}{2}\sum_{i,\alpha\beta}{\vec{\phi}_{i,{\rm SP}}}
{\cdot}C^{\dagger}_{i,\alpha}
{[\mbox{\boldmath{$\vec{\sigma}$}}]}_{\alpha\beta}C_{i,\beta} \nonumber \\
&&\hspace{1cm}\ + \ \ 
\vec{h}\cdot C^{\dagger}_{i,\alpha}
{[\mbox{\boldmath{$\vec{\sigma}$}}]}_{\alpha\beta}C_{i,\beta},\label{2-5-1}
\end{eqnarray}
where $[\mbox{\boldmath{${\sigma}$}}_{\nu}]$ $(\nu = x,y,z)$ 
are the 2$\times$2  Pauli matrices.
This saddle-point approximation is nothing but the mean field
theory and we identify this saddle point solution 
$\vec{\phi}_{i,{\rm SP}}$ with the spin ordering field. 
 We can control this spin ordering field by 
changing $\vec{h}$,  Since the saddle-point 
solution is determined as a function of an 
applied magnetic field ($\vec{h}$). We study 
the MEP induced by an adiabatic deformation of 
this spin ordering field $\vec{\phi}_{i,{\rm SP}}$, 
which we will call $\vec{\varphi}$ in the followings.

When we consider a commensurate magnetic order, the wavefunction
of an electronic state is given by the Slater determinant
composed by the magnetic Bloch wavefunctions; 
\begin{eqnarray} 
\langle \vec{R}_l ,a,\mu,\alpha\|\Phi_{n,\vec{k},\vec{\varphi}
}\rangle= e^{i\vec{k}\cdot\vec{R}_l}\langle
a,\mu,\alpha|n\rangle, 
\end{eqnarray}
where $\langle a,\mu,\alpha|n\rangle 
\equiv \langle a,\mu,\alpha|n(\vec{k},\vec{\varphi})\rangle $ denotes 
a periodic part of the magnetic Bloch wavefunction and 
$\vec{R}_{l},a,\mu$ and $\alpha$ represent the magnetic unit cell, 
the sublattice, orbital and spin index, respectively. 
By using this single-particle  magnetic Bloch wavefunction,
the electric linear response given in eq.(\ref{2-5}) reduces
to a more compact form in terms of Thouless-Kohmoto-Nightingale-Nijs (TKNN)
formula\cite{TKNN},
\begin{eqnarray}
&&\frac{\partial P_{{\rm el},\mu}}{\partial \vec{\varphi}}\cdot
\delta \vec{\varphi}=-ie\sum_{n: \rm{V.B.}}\sum_{m: \rm{C.B.}}
\frac{1}{(2\pi)^{d}}\int_{\rm{M.B.Z.}}d\vec{k} \nonumber \\
 &&\times \left(\frac{{\Bigl\langle}n\Bigr|
\frac{\partial H_{\rm M.F.}(\vec{k},\vec{\varphi})}{\partial
k_\mu}\Bigl|m{\Bigr\rangle}\Bigl\langle m\Bigr|
\frac{\partial H_{\rm M.F.}(\vec{k},\vec{\varphi})}
{\partial \vec{\varphi}}\Bigl|n \Bigr\rangle}
{(\epsilon_{m}-\epsilon_{n})^{2}}
- {\rm{c.c.}}\right)\cdot \delta \vec{\varphi} ,\nonumber  \\
&&=- ie\sum_{n: \rm{V.B.}}
\frac{1}{(2\pi)^{d}}\int_{\rm{M.B.Z.}}d\vec{k}
\left({\Bigl\langle} \frac{\partial n}{\partial k_\mu}{\Bigr|}
\frac{\partial n }{\partial \vec{\varphi}}{\Bigl\rangle}
 - {\rm{c.c.}}\right)\cdot\delta \vec{\varphi}
\label{2-7},
\end{eqnarray}
where $H(\vec{k},\vec{\varphi})$ is
the Hamiltonian in the
momentum representation and the periodic part of the Bloch w.f.  
$|n\rangle \equiv |n({\vec k},\vec{\varphi})\rangle$ is its
 eigenvector with an eigenenergy 
$\epsilon_{n}\equiv \epsilon_{n}({\vec k},\vec{\varphi})$.
 The inner product between
$\Bigl|\frac{\partial n}{\partial k_{\mu}}\Bigr\rangle 
\equiv \frac{\partial}{\partial k_\mu}
\left(|n(\vec{k},\vec{\varphi})\rangle\right)$ and
$\Bigl|\frac{\partial n}{\partial \varphi_{\nu}}\Bigr\rangle$ 
denotes the contraction over 
sublattice, orbital and spin indice, i.e. 
$\sum_{a,\mu,\alpha}(\frac{\partial }{\partial k_\mu}
\langle n(\vec{k},\vec{\varphi})|a,\mu,\alpha\rangle)
(\frac{\partial}{\partial \vec{\varphi}}
\langle a,\mu,\alpha |n(\vec{k},\vec{\varphi})\rangle)$.
Here the abbreviation ``C.B.'', ``V.B.'' and ``M.B.Z.'' represent the
conduction bands, the valence bands and the magnetic Brillouin
zone, respectively. 
\begin{figure}[t]
\begin{center}
\includegraphics[width=0.3\textwidth]{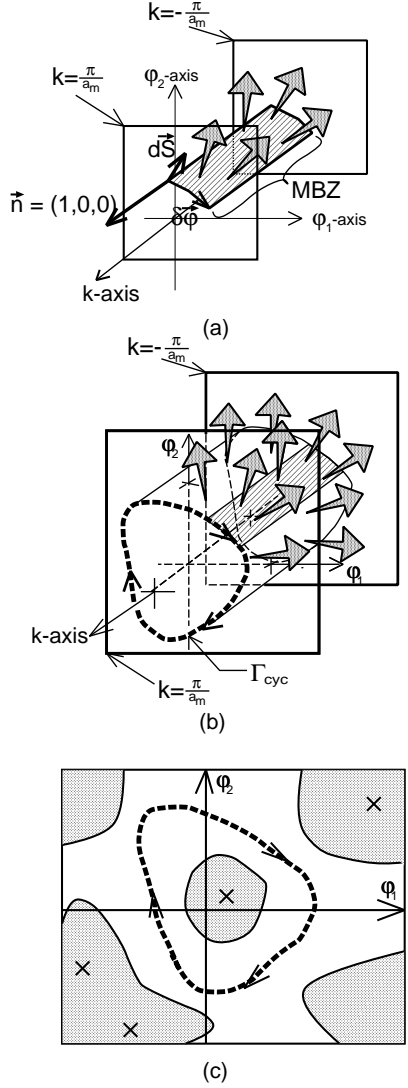}
\end{center}
\caption{(a) A shadowed rectangular area is spanned by $\delta\vec{\varphi}$
and the line parallel to $k$-axis, i.e.,  
$\delta\vec{\varphi}\times
[-\frac{\pi}{\rm{a_m}},\frac{\pi}{\rm{a_m}}]$.
$\vec{n}$, $\delta\vec{\varphi}$ and $d\vec{S}$ form the right hand 
coordinate. The gray arrows represent the flux defined 
in eq.(\ref{2-8}). (b) A cyclic deformation of the external  
parameters $\vec{\varphi}$ in $\varphi_{1}-\varphi_{2}$ space 
is defined by $\Gamma_{\rm cyc}$. 
(c) Shaded regions are unphysical parameter regions (metallic regions)
where a band gap closes, while, the white area denotes a gapped region. 
A cross ($\times$) denotes the parameter point where the H.V.B. and L.C.B. 
forms doubly degenerate point (magnetic monopole). These parameter
 points are included in the metallic regions. The MEP 
induced by the adiabatic change in the gapped region 
(denoted by broken line) is governed by the doubly degenerate 
points located in the metallic region.}
\label{1}
\end{figure}

When we take $\vec{\varphi}$ as a two-component vector, i.e.  
$\vec{\varphi}=(\varphi_{1},\varphi_{2})$,  and consider
the one-dimensional system, we can introduce 
a 3-dimensional {\it generalized momentum space} 
spanned by the crystal momentum $k$ and these 
external parameters ${\varphi}_{1}$ and ${\varphi}_{2}$. In the 
following, the range of the magnetic Brillouin zone (M.B.Z.)
is denoted as $[-\frac{\pi}{{\rm{a_m}}},\frac{\pi}{{\rm{a_m}}}]$, 
where ${\rm{a_m}}$ is a lattice constant of the magnetic unit cell. 
In this generalized momentum 
space, we next introduce the following vector fields for each 
energy band $n$, 
\begin{eqnarray}
\vec{\cal B}_{n}(k,\varphi_{1},\varphi_{2})
&=&\vec{\nabla}\times\vec{\cal A}_{n}(k,\varphi_{1},\varphi_{2}),
\label{2-8} \\
\vec{\cal A}_{n}(k,\varphi_{1},\varphi_{2}) &=& - 
i\langle n(k,\varphi_{1},\varphi_{2})|\vec{\nabla}
|n(k,\varphi_{1},\varphi_{2})\rangle,\label{2-9}
\end{eqnarray}
where $ \vec{\nabla} 
\equiv ({\partial}_{k},{\partial}_{\varphi_{1}},
{\partial}_{\varphi_{2}})$. 
This vector field $\vec{\cal B}_{n}(k,\varphi_{1},\varphi_{2})$
is independent of the following U(1) gauge transformation 
of the magnetic Bloch wavefunction,  
\begin{eqnarray}
&&\langle a,\mu,\alpha
|n,k,\varphi_{1},\varphi_{2}\rangle \nonumber \\ 
&&\rightarrow\  e^{i\xi(k,\varphi_{1},\varphi_{2})}\langle a,\mu,\alpha
|n,k,\varphi_{1},\varphi_{2}\rangle,
\end{eqnarray}
while $\vec{\cal A}_{n}(k,\varphi_{1},\varphi_{2})$
changes by $\vec{\nabla}\xi(k,\varphi_{1},\varphi_{2})$.
Accordingly, we call $\vec{\cal B}_{n}(k,\varphi_{1},\varphi_{2})$
and $\vec{\cal A}_{n}(k,\varphi_{1},\varphi_{2})$  as the
flux and gauge field, respectively. 
By using these vector fields, eq.(\ref{2-7})
can be expressed as a surface integral of the flux
over the rectangular surface spanned by $\delta \vec{\varphi}$ and
the $k$-axis, which is drawn as the shadowed
region in Fig.\ref{1}(a),
\begin{eqnarray}
\frac{\partial P_{\rm{el}}}{\partial \vec{\varphi}}
\cdot \delta {\vec \varphi}
=\frac{e}{2\pi}\sum_{n:{\rm{V.B.}}}\int_{\delta\vec{\varphi}\times
[-\frac{\pi}{{\rm{a_m}}},\frac{\pi}{{\rm{a_{m}}}}]}d\vec{S}\cdot
\vec{\cal B}_{n}(k,\varphi_{1},\varphi_{2}). \label{2-10}
\end{eqnarray}
The direction of $d\vec{S}$ in eq.(\ref{2-10}) is taken so that
$\vec{n}-\delta\vec{\varphi}-d\vec{S}$
forms the right-hand coordinate where $\vec{n}$ represents
a unit vector along the $k$-axis, i.e., $\vec{n}=(1,0,0)$
as in Fig.\ref{1}(a). 

Our strategy is first (i) to reveal the distribution 
of the flux for every filled energy band and then (ii)  
to determine the {\it{optimal}} direction, 
$\delta\vec{\varphi}_{\rm{opt}}$, which maximizes the 
magnetoelectric response given in eq.(\ref{2-10}).
For this purpose, we need to identify 
the {\it{source}} and/or {\it{sink}} of the flux, 
which we will discuss in the next subsection.
\subsection{Dirac monopole and Quantum Pump}
In the region where 
the flux is well-defined, i.e. the $n$-th band 
is isolated from its neighboring bands, 
 there are neither source nor sink associated with 
 the flux $\vec{B}_{n}$, because 
$\vec{\nabla}\cdot\vec{B}_{n}=
\vec{\nabla}\cdot(\vec{\nabla}\times\vec{A}_{n})=0$.
 However in those points where the the $n$-th band 
forms a {\it degeneracy} with its neighboring bands, 
the flux  $\vec{B}_{n}$ becomes ill-defined. Therefore,  
these degeneracy points might become 
sources and/or sinks of the flux. 

As shown in Appendix B, 
when the $n$-th band and its neighboring 
band, e.g.$(n+1)$-th band, are degenerate 
at $(k,{\varphi}_{1},{\varphi}_{2})=
(k_{\rm{D}},{\varphi}_{1,\rm{D}}
,{\varphi}_{2,\rm{D}})$,  this degenerate 
point becomes in general source and/or sink 
for the flux $\vec{B}_{n}$ (and also $\vec{B}_{n+1}$) 
with charge $2\pi$. 
We will call these doubly degenerate points as 
magnetic (anti-)monopoles or Dirac (anti-)monopoles
from now on. Near this doubly degenerate point, 
the effective Hamiltonian for these two bands
can be in general expanded as follows ,
\begin{eqnarray}
&&\mbox{\boldmath{$[H]$}}_{2\times2}(k,\varphi_{1},\varphi_{2})
\nonumber \\
&&\hspace{0.3cm} = 
(\epsilon_{n}(k_{\rm{D}},\varphi_{1,\rm{D}},\varphi_{2,\rm{D}})
+ \vec{K}\cdot\vec{c})\mbox{\boldmath{$[1]$}} \nonumber \\
&&\hspace{0.7cm} +  \ \sum_{\mu,\nu}K_{\mu}V_{\nu,\mu}
[{\mbox{\boldmath{$\sigma$}}}_{\nu}],\label{2-12}
\end{eqnarray}
where $\vec{K}$ is $(k-k_{\rm{D}},\varphi_{1}-\varphi_{1,\rm{D}}
,\varphi_{2}-\varphi_{2,\rm{D}})$, 
and $[{\mbox{\boldmath{$\sigma$}}}_{\nu}]$
are the 2$\times$2  Pauli matrices. The vector $\vec{c}$ is not important
for our purpose, while the determinant of the 3$\times$3
matrix $V$ plays an important role in classifying
this degeneracy point. According to Appendix B, 
for the flux of the upper band ,i.e. $\vec{\cal{B}}_{n+1}
(k,\vec{\varphi})$, this degeneracy point acts as an
anti-monopole (sink) in the case of ${\rm{det}}V<0$,
while as a monopole (source) in the case of ${\rm{det}}V>0$. 
This relation is reversed for the flux of the
lower band $\vec{\cal{B}}_{n}(k,\vec{\varphi})$, i.e.  
\begin{eqnarray}
\vec{\nabla}\cdot \vec{\cal{B}}_{n} &=&
-2\pi\cdot{\rm{sign}}({\rm{det}}V)\delta^{3}(\vec{K}),\label{2-11-1} \\
\vec{\nabla}\cdot \vec{\cal{B}}_{n+1} &=&
2\pi\cdot{\rm{sign}}({\rm{det}}V)\delta^{3}(\vec{K}).\label{2-11-2}
\end{eqnarray}

Higher order degenerate points such as fourth-fold 
degenerate points have also a chance to 
become sources and/or sinks for the flux $\vec{B}_{n}$. 
However it often happens that these higher order degenerate 
points in the 3D parameter space can be decomposed into 
several number of doubly degenerate points, namely, an effective 
Hamiltonian reduces to the direct sum of two by two 
matrices given in eq.(\ref{2-12}). Therefore we 
will concentrate on the doubly degenerate points in the following.    

Due to the nature of doubly degenerate points given 
in eqs.(\ref{2-11-1}) and (\ref{2-11-2}), when we fill both 
of these two bands by electrons, 
$\vec{\cal{B}}_{n}(k,\vec{\varphi})$ and
$\vec{\cal{B}}_{n+1}(k,\vec{\varphi})$ cancel each 
other in the summation of the 
r.h.s. of eq.(\ref{2-10}) and 
the degeneracy point between the $n$-th and $(n+1)$-th 
band becomes almost irrelevant for the MEP. 
On the other hand, when only the lower band is filled by electrons,
$\vec{\cal{B}}_{n}(k,\vec{\varphi})$ contributes to the MEP without being
cancelled by its counter part $\vec{\cal{B}}_{n+1}(k,\vec{\varphi})$
and results in a gigantic linear response, where 
the doubly degenerate point between these two bands 
plays an essential role in the direction and magnitude of the MEP.
Those magnetic dielectrics which belong to the latter case of 
the electron filling are good candidates where 
 this mechanism becomes relevant. Based on 
a specific model, we will discuss in section 3 
about the physical consequences of 
these two types of electron fillings.  

When we fix the electron number per site, those parameter 
points at which the highest valence band (H.V.B.) 
and the lowest conduction band (L.C.B.) form doubly degenerate 
points are usually unstable points (unphysical points). This 
is because making a direct band gap lowers the energy 
of a ground state wavefunction in general. Then the 
spin ordering field usually takes those parameter 
regions where these degeneracies are lifted. 
We will argue later that this is indeed the case 
with a specific model given in section 3. 
What is important and nontrivial is 
that {\it the MEP induced by the adiabatic change in these 
gapped phase  is still controlled by the doubly degenerate 
points hidden in its neighboring unphysical parameter 
points} (See Fig.\ref{1}(c)). 

When we deform $\vec{\varphi}$ by a finite amount and
put it back to the initial value $\vec{\varphi}_{i}$, i.e.,
making a loop $\Gamma_{\rm{cyc}}$ in the $\varphi_{1}-\varphi_{2}$
plane as in Fig.\ref{1}(b),
the total change of the electronic polarization induced by this process
is the total flux penetrating the cylinder surface
spanned by $\Gamma_{\rm{cyc}}$ and $k$-axis, i.e., $\Gamma_{\rm{cyc}}\times
[-\frac{\pi}{{\rm{a_{m}}}},\frac{\pi}{{\rm{a_m}}}]$,
\begin{eqnarray}
\Delta_{\rm{cyc}}P_{\rm{el}}&\equiv&
\oint_{\Gamma_{\rm{cyc}}}
\frac{\partial P_{\rm{el}}}{\partial \vec{\varphi}}
\cdot d{\vec \varphi},\nonumber \\
&=&\frac{e}{2\pi}\sum_{n:{\rm{V.B.}}}\int_{\Gamma_{\rm{cyc}}\times
[-\frac{\pi}{{\rm{a}_{m}}},\frac{\pi}{{\rm{a}_{m}}}]}d\vec{S}\cdot
\vec{{\cal{B}}_{n}}.\label{2-10-1} 
\end{eqnarray}
This is because the surface integrals of the flux $\vec{\cal{B}}_{n}$
over the $k=\frac{\pi}{\rm{a}_{m}}$ plane and that
over the $k=-\frac{\pi}{\rm{a}_{m}}$ plane cancel 
each other due to the periodicity. Therefore,  
if this cylinder has some magnetic monopoles and/or 
antimonopoles of $\vec{\cal B}_{n}$ inside, 
we have a chance to obtain a non-zero change
in the MEP after this cyclic deformation 
along $\Gamma_{\rm{cyc}}$.
This kind of the ``d.c.'' response induced by the ``a.c.'' impulse 
is called Quantum Pump\cite{thouless,altshuler}, 
since finite amounts of electrons  
are transported and/or {\it{pumped}} up from the one side 
to the other side of the system after one cycle of this adiabatic 
deformation.

By using eq.(\ref{2-11-1}) and eq.(\ref{2-11-2}), we can  
simplifies eq.(\ref{2-10-1}) into more intuitive form. 
Namely, when both the $(n+1)$-th and $n$-th bands are filled,
a pair of magnetic monopole(anti-monopole) for the $(n+1)$-th band
and  anti-monopole(monopole) for the $n$-th band cancel each other
in the summation of the r.h.s. of eq.(\ref{2-10-1}).
As a result of this cancellation, the degeneracy points 
between {\it{filled}} bands do not contribute to 
the $\Delta_{\rm{cyc}}P_{\rm{el}}$ in eq.(\ref{2-10-1}). 
Then a finite $\Delta_{\Gamma_{\rm{cyc}}} P_{\rm{el}}$ 
is originated only from 
{\it{the degeneracy points where the highest valence band (H.V.B.) and
the lowest conduction band (L.C.B.) touch}}. Namely, we obtain a 
following simple form for the quantized particle transport:
\begin{eqnarray}
\Delta_{\rm{cyc}}P_{\rm{el}}= -e\times\sum_{i=1}^{N}
{\rm{sign}}({\rm{det}}V_{i}), \label{2-13}
\end{eqnarray}
where the index $i$ represents the 
doubly degenerate point between the H.V.B. and the L.C.B..
$N$ is the total number of these degenerate points. 
The 3$\times$3 matrix $V$ given in eq.(\ref{2-12}) is defined 
for each doubly degenerate point and $V_{i}$ 
denotes that of the $i$-th point. 

If the path $\Gamma_{\rm{cyc}}$ is 
contractible into one point without confronting a metallic region
where the H.V.B. and the L.C.B. form a band crossing, 
the left hand side of eq.(\ref{2-13}) is trivially zero and 
so is $\Delta_{\Gamma_{\rm{cyc}}} P_{\rm{el}}$. 
Therefore eq.(\ref{2-13}) indicates that we must choose 
the path $\Gamma_{\rm{cyc}}$ to  be at least 
{\it noncontractible}, in order to get non-zero 
$\Delta_{\Gamma_{\rm{cyc}}} P_{\rm{el}}$.

Finally but not the least, we want to mention about the 
stability of the magnetic (anti-)monopole. 
As is evident from eq.(\ref{2-12}), 
we need to tune  only 3 {\it{real}}-valued parameters 
in order to lift the double degeneracy. 
Therefore, when we add another parameter $\varphi_{{\rm{perturb}}}$
and constitute a four dimensional parameter
space, the double degeneracy occurs on a {\it{line}} in this 4D space.
As a result of this line degeneracy appearing in the 4D parameter 
space, these magnetic (anti)monopoles are 
stable against small perturbations. 
We will also revisit this point in section 3, 
where we will generalize the 1D studies into 
the 2D case by introducing weak interchain 
hopping terms and demonstrate that the magnetic (anti)monopole 
observed in the 1D studies is in fact stable 
against this interchain hoppings.

\section{Gigantic ME Effect  --- spin-orbit coupling model --- }
Based on the concept developed in section 2,  
we will construct a model 
where the {\it{global}} rotation in a spin space 
induces a gigantic electronic polarization. In this model, 
the spin-orbit coupling plays a crucial role, because the 
global rotation in the spin space cannot affects its electronic 
state without spin-orbit interactions.

\begin{figure}[t]
\begin{center}
\includegraphics[width=0.4\textwidth]{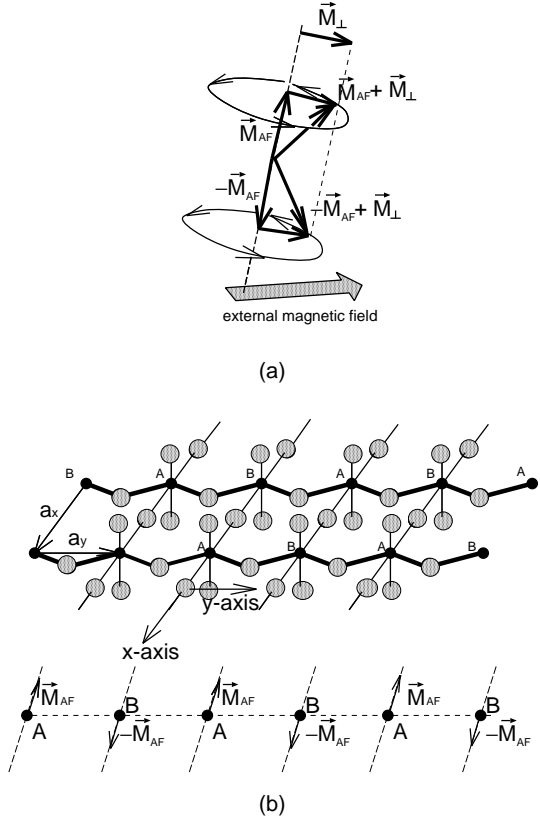}
\end{center}
\caption{(a) The sublattice magnetization  
$\vec{M}_{\rm{AF}}+\vec{M}_{\perp}$ is for the A sublattice, 
while $-\vec{M}_{\rm{AF}}+\vec{M}_{\perp}$ is 
for the B sublattice, where A and B sublattice are defined 
in Fig.(b). The ferromagnetic moment $\vec{M}_{\perp}$ 
can be controlled by an applied magnetic field. 
(b) A modulated square lattice with the staggered magnetization on 
A and B sublattice. Grey circles denote the anion 
atoms 
while the transition metal atoms are located on the lattice. 
Here we draw the 1D chains extending along the $y$-direction
 as bold lines, while the interchain couplings are introduced 
as thin lines. Futhermore, the anion atoms 
on the 1D chain are shifted along the $x$-direction uniformly.}
\label{5}
\end{figure}

Let's consider an antiferromanget on a modulated square lattice, 
where its sublattice magnetizations collinear to a particular direction
($\vec{M}_{\rm AF}$) cant and acquire a ferromagnetic moment  
$\vec{M}_{\perp}$ as shown in Fig.\ref{5}. 
We can control the direction and magnitude of this 
ferromagnetic moment $\vec{M}_{\perp}$ by using an external 
magnetic field.
The hopping integral between the nearest neighbor sites
depends on the intermediate anion ions,
for example, oxygen ${\rm{O^{2-}}}$. In Fig.\ref{5}(b),
we introduce the uniform shifts of the anion atoms along
the $x$-direction, since our spin
structure itself does not break the spatial inversion symmetry.

By adding an orbital index on each site in the tight binding model,
we study the dielectric property of a model composed of 
the $d$-orbitals
with the spin-orbit interaction and the canted antiferromagnetic order.
At first, we give the result of the 1-dimensional system extending
along the $y$-direction (bold line in the upper panel of Fig.\ref{5}(b)). 
Next we give the 2-dimensional generalization  
by adding weak electron transfers along the $x$-direction.
This generalization will be justified, since the 
magnetic (anti)monopoles we will observe in the 1D studies 
are actually stable against small perturbations such as 
an interchain hopping term.

The crystal field term $H_{\rm C.F.}$ and the kinetic energy 
extending along the $y$-direction $H_{{\rm K},y}$ 
are given as follows.
\begin{eqnarray}
H_{\rm CF}&=&
\sum_{i_{y},\mu,\alpha}\epsilon_{\mu}C^{\dagger}_{i_{y},\mu,\alpha}
C_{i_{y},\mu,\alpha},\nonumber \\
H_{{\rm K},y}&=&\sum_{i_{y},\mu,\nu,\alpha}C^{\dagger}_{i_{y}
+{\rm{a}}_{y},\mu,\alpha}
\left[{\mbox{\boldmath{$t$}}}^{y}\right]_{\mu\nu}C_{i_{y},\nu,\alpha}
+ {\rm{c.c.}},\nonumber \\
\left[\mbox{\boldmath{$t$}}^{y}\right]
&=&\left[\mbox{\boldmath{$t$}}^{y,s}\right] +
\left[\mbox{\boldmath{$t$}}^{y,a}\right],\nonumber \\
&=&\left[\begin{array}{ccccc}
-t_{0}&-t_{1}&0&0&0 \\
-t_{1}&-t_{2}&0&0&0 \\
0&0&-t_{5}&0&0 \\
0&0&0&-t_{5}&0 \\
0&0&0&0&0
\end{array}\right]
\nonumber \\ 
&&
+\ \ \left[\begin{array}{ccccc}
0&0&-t_{3}&0&0 \\
0&0&-t_{4}&0&0 \\
t_{3}&t_{4}&0&0&0 \\
0&0&0&0&t_{6} \\
0&0&0&-t_{6}&0
\end{array}\right].\label{3-8}
\end{eqnarray}
Here $\epsilon_{\mu}$ is the energy 
of the $d$-orbital: $\mu = d_{x^2-y^2},d_{3z^2-r^2},d_{xy},d_{yz},d_{zx}$.
The transfer integrals are composed of the symmetric parts
$\left[\mbox{\boldmath{$t$}}^{y,s}\right]$ and
antisymmetric parts $\left[\mbox{\boldmath{$t$}}^{y,a}\right]$.
The antisymmetric transfer integrals $t_{3},t_{4},t_{6}$ result
from the anion ions' shifts and their signs are reversed when the 
anion ions slide in the opposite direction ($-x$ direction). Namely,  
these transfer integrals break the spatial inversion symmetry $I$.
Also $t_{1}$ is induced by the displacement of the anion ions.
An on-site spin-orbit coupling is introduced as follows, 
\begin{eqnarray}
H_{\rm LS}&=&\sum_{\alpha,\beta,\mu,\nu}
\lambda C^{\dagger}_{i,\mu,\alpha}
{\mbox{\boldmath{$[\vec{L}]$}}}_{\mu\nu}{\mbox{\boldmath{$\cdot$}}}
{\mbox{\boldmath{$[\vec{\sigma}]$}}}_{\alpha\beta}
C_{i,\nu,\beta},\nonumber \\
\left[{\mbox{\boldmath{$L$}}}_{x}\right]&=&\left[\begin{array}{ccccc}
0&0&0&-i&0 \\
0&0&0&-\sqrt{3}i&0 \\
0&0&0&0&i \\
i&\sqrt{3}&0&0&0 \\
0&0&-i&0&0
\end{array}\right],\nonumber \\
\left[{\mbox{\boldmath{$L$}}}_{y}\right]&=&\left[\begin{array}{ccccc}
0&0&0&0&i \\
0&0&0&0&-\sqrt{3}i \\
0&0&0&i&0 \\
0&0&-i&0&0 \\
-i&\sqrt{3}i&0&0&0
\end{array}\right],\nonumber \\
\left[{\mbox{\boldmath{$L$}}}_{z}\right]&=&\left[\begin{array}{ccccc}
0&0&2i&0&0 \\
0&0&0&0&0 \\
-2i&0&0&0&0 \\
0&0&0&0&-i \\
0&0&0&i&0
\end{array}\right].\label{3-9}
\end{eqnarray}
As we explained in section 2, our spin ordering field  
, i.e. the canted AF magnetic moment, enters into our 
mean-field Hamiltonian in the following way:
\begin{eqnarray}
H_{\rm MF} = 
\frac{U}{2}\sum_{i,\mu,\alpha,\beta}\vec{\varphi}_{i}
{\cdot}C^{\dagger}_{i,\mu,\alpha}
{\mbox{\boldmath{$[\vec{\sigma}]$}}}_{\alpha\beta}
C_{i,\mu,\beta}.\label{3-10}
\end{eqnarray} 
Here $\vec{\varphi}_{i}$ is taken as 
$(-)\vec{M}_{\rm AF}+\vec{M}_{\perp}$ when $i$ 
belongs to the A(B)-sublattice. The two-component 
ferromagnetic moment $\vec{M}_{\perp}=(M_{\perp,x},M_{\perp,y})$
can be controlled by the applied magnetic field 
as we mentioned above. 

\subsection{Odd number of electrons filling}
\begin{figure}[t]
\begin{center}
\includegraphics[width=0.6\textwidth]{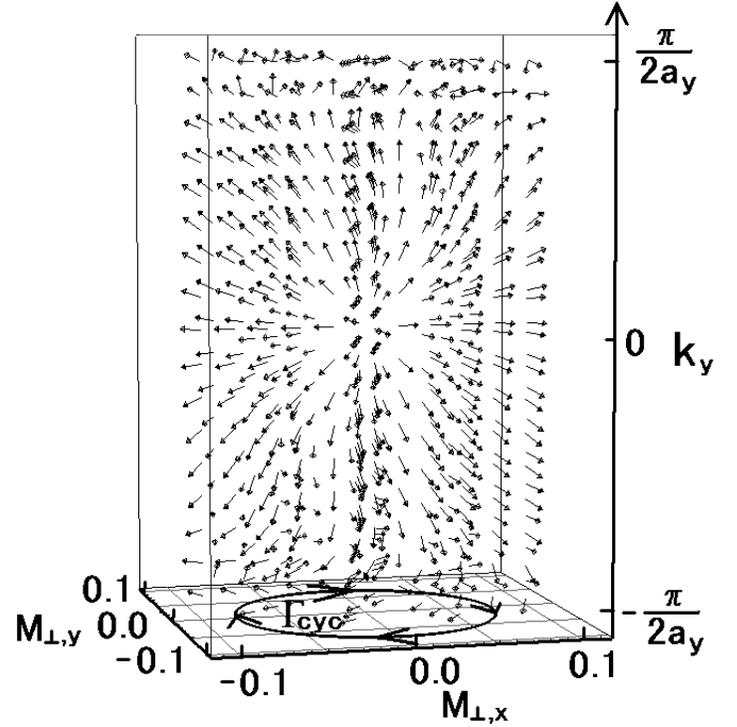}
\end{center}
\caption{A distribution of the flux $\vec{\cal B}$ defined 
in eq.(\ref{3-11-0}). 
$t_{0}/U=0.125$, $t_{1}/U=0.1$, 
$t_{2}/U=0.025$, $t_{3}/U=0.075$, 
$t_{4}/U,t_{5}/U,t_{6}/U=0.05$, 
$\lambda/U=0.025$,  
$\epsilon_{x^2-y^2}/U=1.25$, 
$\epsilon_{3z^2-r^2}/U=1.3$, 
$\epsilon_{xy}/U=0.05$, 
$\epsilon_{yz}/U=0.0$, 
$\epsilon_{zx}/U=0.05$, 
$\vec{M}_{\rm AF}\parallel [1,1,1]$ and 
$|\vec{M}_{\rm AF}|=1$. 
We take the $M_{\perp,x}$-axis and $M_{\perp,y}$-axis as   
$(\cos{\theta}\cos{\phi},\cos{\theta}\sin{\phi},-\sin{\theta})$
and $(-\sin{\phi},\cos{\phi},0)$ respectively, 
where $\vec{M}_{\rm AF}=
(\sin{\theta}\cos{\phi},\sin{\theta}\sin{\phi},\cos{\theta})$.
The vertical axis of this figure is the $k_{y}$-axis 
whose range is taken to be $[-\frac{\pi}{2{\rm{a}}_{y}},
-\frac{\pi}{2{\rm{a}}_{y}}]$ and we take ${\rm{a}}_{y}=10$ for visibility. 
We normalize the vectors to be the same in order to 
show only their directions.}
\label{6}
\end{figure}

\begin{figure}[t]
\begin{center}
\includegraphics[width=0.49\textwidth]{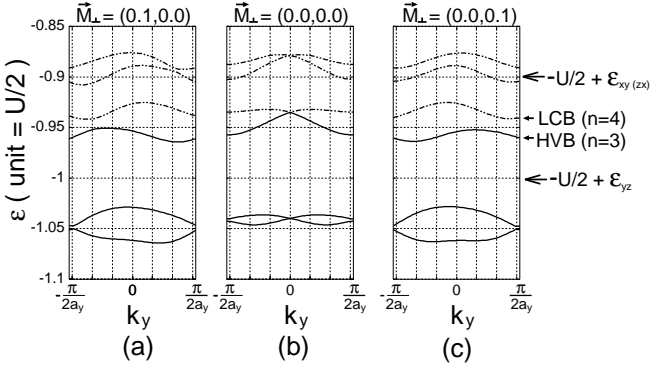}
\end{center}
\caption{Energy dispersions of the lower 6 bands, which are mainly 
composed of $t_{2g}$ orbitals and constitute the lower Hubbard bands. 
Here we take the same parameter 
values as those in Fig.\ref{6}. The broken lines 
denote those of the empty bands, while the 
solid lines represent the filled bands 
(in the case of 3 electrons per magnetic unit cell).} 
\label{7}
\end{figure}

Based on this model Hamiltonian, we consider a situation 
where odd number of electrons are filled per magnetic unit cell. 
To make the following analysis specific, we fill 3 electrons 
per magnetic unit cell. In order to discuss the dielectric property 
of this electron filling, we calculate a following  
flux $\vec{\cal B}$: 
\begin{eqnarray}
\vec{\cal{B}}(\vec{M}_{\perp},k_y)\equiv\sum_{n=1}^{3}
\vec{\cal{B}}_{n}(\vec{M}_{\perp},k_y) \label{3-11-0},  
\end{eqnarray}
where a flux for each energy band, i.e. 
$\vec{\cal{B}}_{n}(\vec{M}_{\perp},k_{y})$, is 
defined in eqs.(\ref{2-8},\ref{2-9}),
where $\vec{\nabla}$  in these equations  
is regarded as $(\frac{\partial}{\partial M_{\perp,x}},
\frac{\partial}{\partial M_{\perp,y}}
,\frac{\partial}{\partial k_{y}})$ in the above equation. 
The crystal momentum $k_y$ ranges 
$[-\frac{\pi}{2{\rm{a}}_{y}},\frac{\pi}{2{\rm{a}}_{y}}]$. 
The distribution of this flux in the 
$\vec{M}_{\perp}-k_{y}$ space is given  
in Fig.\ref{6}, where we find a source of 
the flux, i.e., a magnetic monopole 
at $(\vec{M}_{\perp},k_y)=(0,0,0)$. 
This source corresponds to the doubly degenerate point  
which the H.V.B. ($n=3$) 
and the L.C.B. ($n=4$) 
form at $\Gamma$ point in the case of 
$\vec{M}_{\perp}=0$ 
(see Fig.\ref{7}(b)). This band touching 
point is nothing but the Kramers 
doublet, since the collinear antiferromagnet 
($\vec{M}_{\perp}=0$) is invariant under 
the time-reversal operation $R$ combined with the spatial 
translation by ${\rm a}_y$;  
$\{R| {\rm a}_{y}\}$.  Namely, following two Bloch 
wavefunctions  
are degenerated at $k_y=0$ point\cite{symmetry}:
\begin{eqnarray}
&&\langle a,\mu,\alpha|n,\vec{M}_{\perp}=0,k_y\rangle 
=\nonumber \\
&&\sum_{b,\beta}\left[
\mbox{\boldmath{$\sigma$}}_{x}\right]_{ab}
\left[-i\mbox{\boldmath{$\sigma$}}_{y}
\right]_{\alpha\beta}
\langle b,\mu,\beta|n,\vec{M}_{\perp}=0,-k_y
\rangle^{\ast}\label{3-11}, 
\end{eqnarray}
where $a$ and $b$ denote the sublattice index.
$\left[\mbox{\boldmath{$\sigma$}}_{x}\right]_{ab}$ 
in eq.(\ref{3-11}) exchanges A and B sublattice and thus  
represents the translation by ${\rm a}_y$. 
In the case of our electron-filling (3 electrons per 
magnetic unit cell), 
a finite magnetization $\vec{M}_{\perp}$ always appears 
and lifts this degeneracy as in Fig.\ref{7}(a,c). 
This is because making a band gap lowers the total 
energy of the ground state wavefunction, which we can 
understand by comparing the energy dispersion at $\vec{M}_{\perp}=0$ 
and that of finite $\vec{M}_{\perp}$. 
(compare Fig.\ref{7}(b) and (a,c)). 
Thereby the doubly degenerate point we mentioned above 
turns out to locate at an unstable parameter point.  
However, this degeneracy still plays an important role for 
the dielectric property at {\it finite}-$\vec{M}_{\perp}$, 
as we explained in section 2. Namely, even 
when $\vec{M}_{\perp}$ is deformed within the gapped region 
(finite-$\vec{M}_{\perp}$ region), the induced electronic 
polarization $P_{y,\rm el}$ is determined by the flux 
$\vec{\cal{B}}(\vec{M}_{\perp},k_y)$ whose nontrivial 
distribution is originated from this doubly degenerate point 
at $\vec{M}_{\perp} = 0$.  
Futhermore, when $\vec{M}_{\perp}$ is deformed along the path
$\Gamma_{\rm{cyc}}$ which encloses 
$\vec{M}_{\perp}=0$ point clockwise (Fig.\ref{6}), 
an induced electronic polarization 
$\Delta_{\rm cyc} P_{y,\rm el}$ amounts 
to $+ e$ due to this magnetic monopole. These features 
do not alter drastically as far as we fill odd number of 
electrons per magnetic unit cell. 
Namely, we always find a magnetic monopole 
or antimonopole at this unstable point; 
$\vec{M}_{\perp} = 0$, which gives rise 
to nontrivial dielectric properties in its neighboring  
finite-$\vec{M}_{\perp}$ regions.

\subsection{Even number of electrons filling}
On the contrary, when we fill even number of electrons
 per magnetic unit cell, e.g. 4 electrons per magnetic unit cell, 
the flux which describes the dielectric property, i.e.,  
$\vec{\cal{B}}(\vec{M}_{\perp},k_y)\equiv\sum_{n=1}^{4}
\vec{\cal{B}}_{n}(\vec{M}_{\perp},k_y)$  
has no significant distribution 
in the $\vec{M}_{\perp}-k_y$ space.
This is because the degenerate point at 
$(\vec{M}_{\perp},k_y)=(0,0,0)$ 
also becomes a sink of the flux 
$\vec{\cal{B}}_{4}$. Namely, the direction of  
$\vec{\cal{B}}_{4}$ becomes opposite to that of 
$\vec{\cal{B}}_{3}$ and these two flux strongly 
set off each other. As a result of 
this strong cancellation, the linear 
response $\partial P_{y,\rm el}/
\partial \vec{M}_{\perp}$ itself becomes very 
small in the finite-$\vec{M}_{\perp}$ regions. 
Futhermore, the quantum 
pump does not occur even when $\vec{M}_{\perp}$ is 
deformed around an $\vec{M}=0$ point. This is because, 
at around $\vec{M}=0$, there always remains a finite 
band gap between the H.V.B. $(n=4)$ and L.C.B. $(n=5)$. 

\subsection{2-dimensional generalization}
Let us generalize the above 1D studies into the 2D case 
with 3 electrons per magnetic unit cell. We introduce the interchain 
hopping integrals,  
\begin{eqnarray}
H_{{\rm K},x}&
=&{\bf\Delta}\sum_{\mu,\nu}C^{\dagger}_{i_{x}+{\rm{a}}_{x},i_{y},\mu,\alpha}
\left[\mbox{\boldmath{$t$}}^{x,s}\right]
_{\mu\nu}C_{i_{x},i_{y},\nu,\alpha} 
+\ \  {\rm{c.c.}},\nonumber \\
\left[\mbox{\boldmath{$t$}}^{x,s}\right]
&=&\left[\begin{array}{ccccc}
 -t_{0}&t_{1}&0&0&0 \\
 t_{1}&-t_{2}&0&0&0 \\
  0&0&-t_{5}&0&0 \\
 0&0&0&0&0 \\
 0&0&0&0&-t_{5}
 \end{array}\right].\label{3-12}
\end{eqnarray}
Here we ignore the antisymmetric transfer integrals coming
from the anion ions' shifts,
because they do not change the conclusion essentially as long as  
 they are not so strong. ${\bf{\Delta}}$ represents the anisotropy 
in transfer integrals
between $x$- and  $y$-directions.
Then the $y$-component of electronic polarizations 
$\partial P_{y,{\rm el}}/\partial \vec{M}_{\perp}$ in
eq.(\ref{2-7}) requires another integral with respect to
the crystal momentum $k_{x}$:
\begin{eqnarray}
&&\frac{\partial P_{y,\rm{el}}}{\partial \vec{M}_{\perp}}\cdot\delta
\vec{M}_{\perp}=
\int_{-\frac{\pi}{{\rm{a}}_{x}}}^{\frac{\pi}{{\rm{a}}_{x}}}
\frac{dk_{x}}{2\pi}\frac{\partial P_{y,\rm{el}}(k_{x})}
{\partial \vec{M}_{\perp}}\cdot\delta\vec{M}_{\perp},\label{3-13-1} \\
&&\frac{\partial P_{y,\rm{el}}(k_{x})}
{\partial \vec{M}_{\perp}} = 
\int_{-\frac{\pi}{2{\rm{a}}_{y}}}^{\frac{\pi}{2{\rm{a}}_{y}}} dk_{y}
\sum_{n=1}^{3}\sum_{\mu=x,y}\times \nonumber \\
&&\left(\frac{\partial}{\partial k_{y}}
\langle n(\vec{M}_{\perp},\vec{k})|
\frac{\partial}{\partial M_{\perp,\mu}}
|n(\vec{M}_{\perp},\vec{k})\rangle - {\rm{c.c.}}\right)
\delta M_{\perp,\mu},\nonumber \\
&&=\frac{e}{2\pi}\sum_{n=1}^{3} 
\int_{\delta\vec{M}_{\perp}\times
[-\frac{\pi}{2{\rm{a}}_{y}},\frac{\pi}{2{\rm{a}}_{y}}]}
d\vec{S}\cdot
\vec{\cal B}_{n}(\vec{M}_{\perp},\vec{k}).\label{3-13-2}
\end{eqnarray}

In the followings, we will argue that 
$\partial P_{y,{\rm el}}/\partial \vec{M}_{\perp}$ 
given in eq.(\ref{3-13-1}) will 
exhibit quite similar behaviors as 
that of the previous 1D case, as far as the anisotropy 
parameter ${\bf \Delta}$ is taken to be small.  

When we constitute a four dimensional parameter space spanned 
by $k_x$,$k_y$ and $\vec{M}_{\perp}$,  
the doubly degenerate {\it{point}} we have observed in the 
$\vec{M}_{\perp}-k_y$ space (see Fig.\ref{6}) 
is expected to form a 
degenerate {\it line} in this 4D parameter space. 
This is because only 3 real-valued parameters 
are sufficient to lift a two fold degeneracy 
of an hermite Hamiltonian. 
In the context of our model calculations, 
finite $k_y$, $M_{\perp,x}$ and $M_{\perp,y}$ have 
already lifted the double degeneracy and thus a new 
axis, i.e. $k_x$-axis, cannot lift this degeneracy. In fact, 
a simple analysis given in Appendix C 
proves that the double degeneracy occurs exactly 
on the line $\vec{M}_{\perp}=k_{y}=0$ in this 4D 
parameter space.

As a result of this line degeneracy, the flux for 
every $k_x$, i.e.   
$\sum_{n=1}^{3}\vec{\cal B}_{n}(\vec{M}_{\perp},k_x,k_y)$,  
and its integral over the surface 
$\delta\vec{M}_{\perp}\times
[-\frac{\pi}{2{\rm a}_y},\frac{\pi}{2{\rm a}_y}]$, i.e.  
$\partial P_{y,{\rm el}}(k_x)/\partial 
\vec{M}_{\perp}$ exhibit a similar 
behaviour as that of 
$\sum_{n=1}^{3}\vec{\cal B}_{n}(\vec{M}_{\perp},k_y)$ 
and $\partial P_{{\rm el}}/\partial \vec{M}_{\perp}$ 
in the previous 1D studies.
This is because, for every  $k_{x}=\rm{constant}$ space 
(a 3D space spanned by $k_y$ and $\vec{M}_{\perp}$),
we always find the doubly degenerate point at 
$(\vec{M}_{\perp},k_y)=(0,0,0)$, which becomes a source 
for $\sum_{n=1}^{3}\vec{\cal B}_{n}(\vec{M}_{\perp},k_x,k_y)$.  

\begin{figure}[t]
\begin{center}
\includegraphics[width=0.45\textwidth]{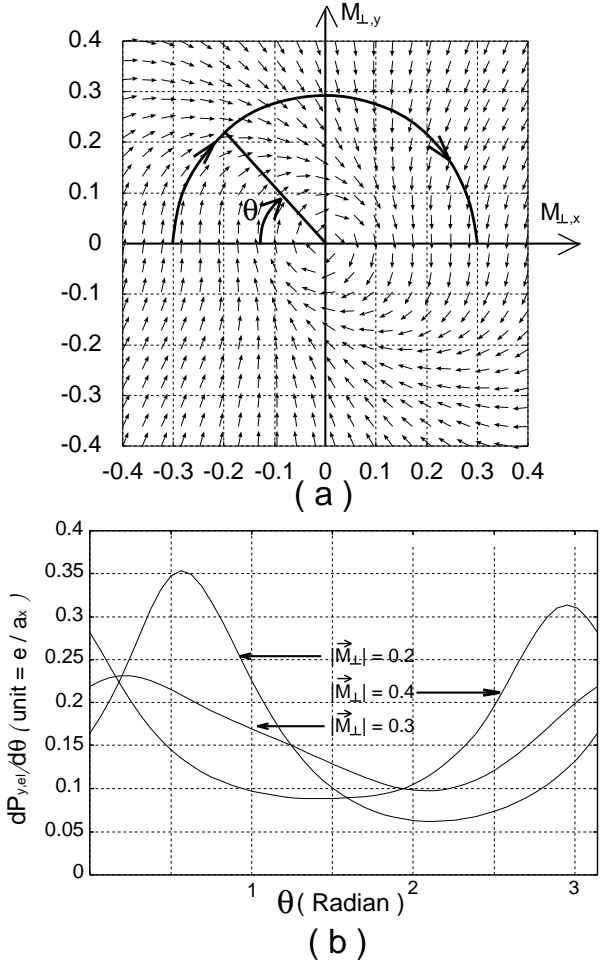}
\end{center}
\caption{(a) The flow of $dP_{y}(\vec{M}_{\perp})/d\vec{M}_{\perp}$
in the 2D case. Here we take ${\bf{\Delta}}=0.2$. The other parameters  
are taken to be same as those of Fig.\ref{6}. 
Here the magnitudes of the vectors are normalized 
to be same, while the informations on their magnitudes 
are partly given in fig.(b). (b) $dP_{y}(|\vec{M}_{\perp}|
,\theta)/d\theta$ as a function of $\theta$, where $\theta$ 
is defined in the text.} 
\label{8}
\end{figure}

After taking the integral of $\partial P_{y,{\rm el}}(k_x)/\partial 
\vec{M}_{\perp}$ over $k_x$, we calculated the electronic 
polarization numerically, which is shown in Fig.\ref{8}.
At the center of the flow diagram shown in Fig.\ref{8}(a), 
we find the vortex, which results from the doubly degenerate 
line we mentioned above. 
In fact, the contour integral of this vector field 
around an arbitrary loop $\Gamma_{\rm{cyc}}$ enclosing this 
vortex clockwise produces $+e/{\rm a}_{x}$, where 
the factor $1/{\rm a}_{x}$ comes from the integral 
over $k_x$ in eq.(\ref{3-13-1}). 

When it comes to the magnitude of $\partial P_{y,{\rm el}}/\partial
\vec{M}_{\perp}$, it becomes larger when the direct energy gap 
becomes smaller. For example, we show the behavior of 
$\partial P_{y,{\rm el}}(|\vec{M}_{\perp}|,\theta)/\partial \theta$ 
as a function of $\theta$ for several constant $|\vec{M}_{\perp}|$ 
in Fig.\ref{8}(b),  where $\theta$ is defined as 
$(M_{\perp,x},M_{\perp,y})=|\vec{M}_{\perp}|
(-\cos\theta,\sin\theta)$. In this figure, we can see that 
$\partial P_{y,{\rm el}}(|\vec{M}_{\perp}|,\theta)/\partial \theta$
enhances around $\theta=0.5\sim 0.6$ in the case 
of $|\vec{M}_{\perp}|=0.2$. This enhancement is 
actually accompanied by the reduction of the direct
band gap. 

When we stack this 2D system in the $z$-direction 
with weak interlayer couplings, the magnitude 
of $\partial P_{y,{\rm el}}/\partial \theta$ amounts to 
$1\ [{\rm{C/m^{2}}}]$ 
in the case of ${\rm a}=4{\rm{\AA}}$:
\begin{eqnarray}
\frac{\partial P_{y,{\rm el}}}{\partial \theta} \simeq  
\frac{e}{{\rm a}^2} \label{3-13-3}.  
\end{eqnarray} 
Meanwhile the electronic energy gain in the presence of the 
finite ferromagnetic moment $\vec{M}_{\perp}$ (see Fig.\ref{7}) is 
phenomenologically ascribed to the symmetric part 
of an anisotropic exchange interactions such as 
$\sum_{i,\mu=x,y,z}\Delta J_{\mu} S_{i,\mu} S_{i+1,\mu}$.\cite{anote1}  
Thus, as far as the linear response is concerned,  
 we can roughly estimate the magnetic field required to rotate 
the spin ordering field as follows: 
\begin{eqnarray}
\frac{\partial \theta}{\partial h} \simeq \frac{g\mu_{\rm B}}{|\Delta J|}. 
\label{3-13-4}
\end{eqnarray}
Accordingly, in combination with eq.(\ref{3-13-3}), 
the {\it magneto-electric coefficient} 
in our model turns out to amount to $10^{-3}[{\rm C/m^2}\cdot {\rm kOe}]$ 
even in the case of $|\Delta J| = 10 {\rm meV}$. 
This value is in fact much larger 
that the typical value $\simeq 2\times 10^{-6}[{\rm C/m^2}\cdot {\rm kOe}]$    
of the conventional magneto-electric materials. On the other 
hand, when the applied magnetic field is sufficiently large 
compared with this anisotropic exchange interactions \cite{anote2} 
($g\mu_{\rm B}h \gg |\Delta J|$), our canted component 
$\vec{M}_{\perp}$ follows the direction of the 
applied field,  namely $\vec{M}_{\perp}\ //\ \vec{h}$.

Finally, we want to mention about the relation 
between the dielectric property 
in our models and the conventional magneto-electric(ME) effect.  
In the conventional ME materials, breaking 
the $IR$ symmetry by applied magnetic fields 
induce a finite electric polarization. In our case, 
on the other hand, the electric polarization is allowed 
over the entire region in Fig.\ref{8}(a). Namely, 
our model with the anion atoms shifted
uniformly already breaks both the spatial inversion symmetry $I$ and its
combination with the time-reversal symmetry $IR$.
However we might call this phenomenon ME 
effect in a wider sense, since 
an applied magnetic field produces an interesting
electric response where a cyclic deformation of the 
background spin configuration 
pumps up electrons toward a particular direction.  

\section{depolarization fields - mean-field arguments - }

Before rushing to concluding remarks, we want to think over 
the effect of depolarization fields on our new mechanism of 
gigantic Magneto-Electric (ME) effect. 
Namely, throughout this article, we have been assuming 
implicitly that our system is embedded into 
a closed circuit, where electronic  
charges transported from one end to the other  
are always short-circuited. However, when the system is {\it terminated 
without electrodes}, the electrons accumulating at one  
boundary cause an electro-static potential inside the system,  
which seems to push back valence band electrons into the system. 
Speaking more generally, when an electric polarization is not 
short-circuited, polarized charges inside the system always 
induce an electric field in the direction {\it reverse} to this  
electric polarization. Thereby one might naturally expect that 
this {\it depolarization} field strongly reduces the Berry phase 
contributions of magneto-electric responses 
we have discussed so far\cite{note6}. 
In this section, we will argue how the magneto-electric response given in 
eq. (\ref{2-7},\ref{2-10}) are suppressed 
in the presence of depolarization field, 
by taking it into account at the mean-field level. 
 
A depolarization field induced by polarized charges 
is widely known to depend on a shape of the system and is 
sometimes complicate to figure out for an arbitrary shape.\cite{LLEDCM} 
Thus, in order to understand as simple as possible 
how the depolarization field affects our novel 
mechanism of ME effect,  
we will take our system to have the most 
simplest geometry. Namely our system is translationally 
invariant along $y$ and $z$ directions and have only 
two plane boundaries at $x=\pm \frac{L}{2}$. 
Then we can take it for granted that both an electric 
polarization and the depolarization field induced by polarized charges    
are always parallel to the $x$-direction. This makes the following 
arguments simple. For example, an electric field penetrating the 
plane $x^{\prime}=x$, which we will call as $E_{d}(x)$, 
 consists of two parts. One is proportional to 
an areal charge density within the region $x\leq x^{\prime}\leq L/2$ and the 
other is to that of $-L/2\leq x^{\prime}\leq x$:  
\begin{eqnarray}
E_{d}(x,\vec{\varphi})= - 
\frac{Q^{+}(x,\vec{\varphi})}{2\epsilon_{0}} +  
\frac{Q^{-}(x,\vec{\varphi})}{2\epsilon_{0}}. \label{2-14-0}
\end{eqnarray}   
In the above equation, we have introduced the areal charge density 
$Q^{\pm}(x,\vec{\varphi})$ as follows:
\begin{eqnarray}
&&Q^{+(-)}(x,\vec{\varphi})\equiv \nonumber \\
&&\frac{1}{S}
\int\!\!\!\int\!\!\!
\int_{x\leq x^{\prime} \leq \frac{L}{2}
\bigl(-\frac{L}{2}\leq x^{\prime}\leq x\bigr)}  
\langle \Psi_{0}(\vec{\varphi})|N_{\vec{r}^{\prime}}| 
\Psi_{0}(\vec{\varphi}) \rangle \ d\vec{r}^{\ \prime},  
\end{eqnarray} 
where $N_{\vec{r}}$ denotes the density operator 
at $\vec{r}=(x^{\prime},y^{\prime},z^{\prime})$ and 
$S$ is the total area of our plane boundary. 

By including the electro-static potential due to this 
depolarization field, we can generalize  
our previous mean field Hamiltonian $H_{\rm M.F.}$, 
namely eq.(\ref{2-5-1}), into the following form: 
\begin{eqnarray}
\bar{H}_{\rm M.F.}(\vec{\varphi})=
H_{\rm M.F.}(\vec{\varphi}) + 
e \sum_{j=1}^{N} E(x_{j},\vec{\varphi})\cdot x_{j},    
\end{eqnarray} 
where $x_{j}$ denotes the $x$-component of a position operator 
of the $j$-th electron.
Correspondingly, the previous expression for the  
variation of the electronic polarization 
 w.r.t. $\vec{\varphi}$ , i.e. eq.(\ref{2-4}), is  
modified into the following equation;
\begin{eqnarray}
\delta P_{{\rm el}} 
= -\frac{i}{V}\sum_{m\ne0}\left[
\frac{\langle\Psi_{0}|J_{\rm el}
|\Psi_{m}\rangle\langle\Psi_{m}|
\delta H_{\rm M.F.}|\Psi_{0}\rangle}
{(E_{m}-E_{0})^{2}} - {\rm{c.c.}}\right]   \nonumber \\ 
 -\ \frac{i e}{V}\sum_{m\ne0}\left[
\frac{\langle\Psi_{0}|J_{\rm el}
|\Psi_{m}\rangle\langle\Psi_{m}|\sum_{j=1}^{N}
\delta E_{d}(x_{j})\cdot x_{j}|\Psi_{0}\rangle}
{(E_{m}-E_{0})^{2}}- {\rm{c.c.}}\right]. \label{2-14}
\end{eqnarray}     
The first term gives the Berry phase contribution of magneto-electric 
effect given in eq.(\ref{2-7},\ref{2-10}), while the latter term brings about 
a negative feedback effect on this topological contributions, which 
we will argue in the followings.   
 
From eq.(\ref{2-14-0}), we can easily relate $\delta E_{d}(x)$ 
appearing in the latter term of eq.(\ref{2-14}) with  
a total electronic current which passes through 
an arbitrary unit area at $x^{\prime}=x$ while the   
perturbation $\delta \vec{\varphi}$ is adiabatically introduced 
\cite{note1}:
\begin{eqnarray}
\delta E_{d}(x) &=& - \frac{1}{\epsilon_{0}} 
\Bigl(Q^{+}(x,\vec{\varphi}+\delta \vec{\varphi}) - Q^{+} 
(x,\vec{\varphi})\Bigr)  \nonumber \\
&\equiv&  -\frac{\bar{J}_{\rm el}(x)}{\epsilon_{0}}. \label{2-15}
\end{eqnarray}
Namely, a total electronic charge 
$Q^{+}(x) + Q^{-}(x)$ is now a conserved quantity in absence of 
electrodes.
This current $\bar{J}_{\rm el}(x)$ is always periodic w.r.t. 
a magnetic primitive translation vector; 
$\bar{J}_{\rm el}(x) = \bar{J}_{\rm el}(x + {\rm a}_{\rm m})$,  
since we have supposed that the perturbation $\delta \vec{\varphi}$ 
does not break the periodicity of the magnetic 
unit cell.\cite{note2}  However its distribution 
{\it within a magnetic unit cell} is not necessarily uniform 
in general, which would make following arguments cumbersome. 
Thus, we assume, for sake of simplicity, that its distribution 
is almost uniform within the magnetic unit cell and replace the 
r.h.s. of eq.(\ref{2-15}) by an electron current integrated 
over the $x$-coordinate \cite{note2};
\begin{eqnarray}
\delta E_{d} (x) \approx \delta E_{d} \equiv - \frac{1}{\epsilon_{0}L}\ 
\int_{-L/2}^{L/2}\bar{J}_{\rm el}(x) \ dx . 
\end{eqnarray}  
Since this integrated charge current is nothing but the change 
of an electronic polarization $\delta \vec{P}_{\rm el}$\cite{resta} 
(see also appendix A), 
 an electric field due to polarized charges now turns out to be 
proportional only to an electronic polarization;
\begin{eqnarray} 
\delta E_{d} = - \frac{1}{\epsilon_{0}} \delta P_{\rm el}.  
\end{eqnarray}

Admitting this relation between an electronic polarization and 
its resulting depolarization 
field $\delta E_{d}(x)$, we could rewrite 
eq. (\ref{2-14}) into a following 
self-consistent equation for $\delta P_{\rm el}$\cite{note2-1,note3};
\begin{eqnarray}
\delta P_{\rm el} 
=  \frac{e}{2\pi}\sum_{n:{\rm{V.B.}}}\int_{\delta\vec{\varphi}\times
[-\frac{\pi}{{\rm{a_m}}},\frac{\pi}{{\rm{a_{m}}}}]}d\vec{S}\cdot
\vec{\cal B}_{n}(k,\varphi_{1},\varphi_{2})
- \chi\ \delta P_{\rm el}. \label{2-16}
\end{eqnarray}     
Here we have introduced the electric susceptibility $\chi$ as   
\begin{eqnarray}
\chi&=& - \frac{i e }{\epsilon_{0} V}\sum_{m\ne0}\left[
\frac{\langle\Psi_{0}|J_{{\rm el}}
|\Psi_{m}\rangle\langle\Psi_{m}|X_{\rm el}|\Psi_{0}\rangle}
{(E_{m}-E_{0})^{2}}- {\rm{c.c.}}\right], \nonumber \\ 
X_{\rm el}&\equiv& \sum_{j=1}^{N} x_{j}. \nonumber 
\end{eqnarray}    
The first term of the r.h.s. in eq.(\ref{2-16}) corresponds to the 
Berry phase contributions to the magneto-electric response, 
while the latter term gives a factor $\frac{1}{1+\chi}$ to 
this topological contributions.  
Namely, the above mean-field equation for $\delta P_{\rm el}$ 
 gives us a following expression for the magneto-electric responses 
in absence of electrodes:    
\begin{eqnarray}
&&\frac{\partial P_{\rm el}}{\partial \vec{\varphi}}
\cdot \delta \vec{\varphi} 
= \nonumber \\
&&\frac{1}{1+\chi}\  
\frac{e}{2\pi}\sum_{n:{\rm{V.B.}}}
\int_{\delta\vec{\varphi}\times
[-\frac{\pi}{{\rm{a_m}}},\frac{\pi}{{\rm{a_{m}}}}]}d\vec{S}\cdot
\vec{\cal B}_{n}(k,\varphi_{1},\varphi_{2}) \label{2-17}. 
\end{eqnarray} 
The electric susceptibility $\chi$ in dielectrics ranges 
from 5 to 5000. Therefore, the magneto-electric coefficient 
we estimated in the previous section (see just below eq.(29)) 
decreases by factor $10^{-1}$ to $10^{-3}$ in presence of 
dipolarization fields.
Futhermore the electric susceptibility 
 usually becomes larger when a direct band 
gap smaller. Thus, in the systems terminated without electrodes, 
our Berry phase mechanism of gigantic ME effect
seems to inevitably meet a severe reduction  
due to this depolarization field. Namely, 
when one look for those mean fields $\vec{\varphi}$ near 
which the H.V.B. and the L.C.B. form magnetic (anti)monopoles, 
not only the Berry phase contributions, i.e. 
$\sum_{n:{\rm{V.B.}}}\vec{\cal B}_{n}(k,\varphi_{1},\varphi_{2})$,  
but the  negative feedback factor $1+\chi$ also 
enhances.
However, when the system in terminated {\it with} electodes and 
electronic polarizations are measured through (short-circuited)
integrated currents, the system is in general free from the 
depolarization field mentioned in this section and thereby 
our Berry phase mechanism of gigantic ME effect 
does not suffer from its negative feedback factor.

\section{Conclusion}
In this article, we have studied the new mechanism
for a gigantic Magneto-Electric (ME) effect.
This is due to the geometrical Berry phase of the Bloch electrons,
which is represented by the flux in the generalized momentum space
including the external parameters such as the direction of the sublattice
magnetization. We proposed a specific model, the multi-band models 
with the spin-orbit interaction, where a magnetic monopole and an
anti-monopole appear in this generalized 
momentum space and determine the distribution of the flux.
These (anti)monopoles are attributed to the Kramers doublet, where the 
broken time-reversal symmetry in addition to the broken spatial 
inversion symmetry is essential.  This model calculation 
indicates that the ME effect is not simply determined by the 
magnitude of the energy gap, and the geometrical structure
 of the U(1) gauge associated with the Bloch
 wavefunctions also plays an important role.  
A comparison with the conventional ME effect
\cite{rado1} suggests that the magnitude could be $ \sim 10^3$ times
larger when the monopole's contribution to the electronic
polarization survives.
Although we do not have an explicit candidate for real systems,
it would be very interesting to look for the material where this
mechanism of gigantic ME effect works. The detailed band structure
calculations will be helpful to reveal the distribution of
the flux in the generalized momentum space, which are
left for future studies.
\begin{acknowledgments}
The authors acknowledge S. Ishihara, Z.Fang, T.Egami S.Murakami,
M.Onoda, and  Y.Tokura for fruitful discussions.
This work is supported by Grant-in-Aids from the Ministry of Education,
Culture, Sports, Science, and Technology of Japan.
\end{acknowledgments}

\appendix
\section{the integrated electron current}

In this Appendix, we will argue that the change 
of the electronic polarization $\delta P_{{\rm el},\mu}$
given in eq.(\ref{2-4}) is identical to a total charge 
transport $\bar{J}_{{\rm el},\mu}$\cite{resta}. This quantity  
is defined as the total current flowing through the system 
while the perturbation $\delta H$ is adiabatically introduced 
from $t=-\infty$ to $t=0$:
\begin{eqnarray}
\bar{J}_{{\rm el},\mu}&\equiv&\frac{1}{V}\int_{-\infty}^{0}
\langle \Psi(t)|J_{{\rm el},\mu}|\Psi(t) \rangle dt 
\label{2-0}. \\
J_{{\rm el},\mu} &\equiv& \int_{V} J_{{\rm el},\mu}(\vec{x}) d\vec{x} 
\nonumber 
\end{eqnarray}
Here $|\Psi(t)\rangle$ is an adiabatically evolved ground state 
wavefunction:
\begin{eqnarray}
i\frac{\partial}{\partial t}|\Psi(t)\rangle = 
\left[\hat{H} + \frac{1}{2}\left(\delta \hat{H}e^{i(\omega - i\delta)t}
+ {\rm H.c.}\right)\right] |\Psi(t)\rangle.\label{2-0-01}
\end{eqnarray}
Therefore, in the translationally invariant systems, 
$\bar{J}_{{\rm el},\mu}$ represents the total charges which pass through 
a unit area normal to the $\mu=$constant plane 
while $t\in [-\infty,0]$. According to the standard time-dependent
perturbation theory, $|\Psi(t)\rangle$ can be 
expanded by the eigenstate $|\Psi_{m}\rangle$ at $t=-\infty$:
\begin{eqnarray} 
&& |\Psi(t)\rangle=|\Psi_{0}\rangle e^{-iE_{0}t}
+ \sum_{m\ne0}|\Psi_{m}\rangle a_{m}(s)e^{-iE_{m}t},\nonumber 
\end{eqnarray}
where its coefficient $a_{m}(t)$ reads: 
\begin{eqnarray}
&&a_{m}(t)= -\frac{e^{i(E_{m}-E_{0}+\omega-i\delta)s}}
{E_{m}-E_{0}+\omega-i\delta}
\langle \Psi_{m}|\frac{1}{2}\delta \hat{H}
|\Psi_{0} \rangle \nonumber \\
&&\hspace{0.8cm}-\frac{e^{i(E_{m}-E_{0}-\omega-i\delta)t}}
{E_{m}-E_{0}-\omega-i\delta}
\langle \Psi_{m}|\frac{1}{2}\delta \hat{H}
|\Psi_{0} \rangle.\nonumber
\end{eqnarray}
Then the electronic current density is 
given up to the first order in $\delta \hat{H}$ as
\begin{eqnarray}
&&\frac{1}{V}\langle \Psi(t)|\hat{J}_{{\rm el},\mu}|\Psi(t) \rangle 
= j_{{\rm el},\mu}(\omega)
e^{i(\omega - i\delta)t} + {\rm{c.c.}},\label{2-0-1} \\
&&j_{{\rm el},\mu}(\omega)\equiv -\frac{1}{V}\sum_{m\ne0}\Bigl(
\frac{\langle\Psi_{0}|
\hat{J}_{{\rm el},\mu}|\Psi_{m}\rangle\langle\Psi_{m}|
\frac{1}{2}\delta \hat{H}|\Psi_{0}\rangle}{E_{m}-E_{0}+\omega -i\delta}
\nonumber \\
&&\hspace{1.5cm}+
\frac{\langle\Psi_{m}|
\hat{J}_{{\rm el},\mu}|\Psi_{0}\rangle\langle
\Psi_{0}
|\frac{1}{2}\delta \hat{H}|\Psi_{m}\rangle}
{E_{m}-E_{0}-(\omega-i\delta)}
\Bigr),\label{2-0-2}
\end{eqnarray}
where we omitted $\langle \Psi_{0}|\hat{J}_{{\rm el},\mu}|\Psi_{0} \rangle$,
because the system is an insulator at $t=-\infty$.
Then by substituting eq.(\ref{2-0-1}) into eq.(\ref{2-0}), 
we obtain $\bar{J}_{{\rm el},\mu}$ in the following way:
\begin{eqnarray}
\bar{J}_{{\rm el},\mu}=
\frac{1}{i(\omega - i\delta)}(j_{{\rm el},\mu}(\omega) 
- j_{{\rm el},\mu}(\omega=0)) + {\rm{c.c.}}.\label{2-0-3}  
\end{eqnarray}
Here we subtracted $2\pi\delta(\omega)j_{{\rm el},\mu}(\omega=0)$ 
from $\bar{J}_{{\rm el},\mu}$, because we assume that the system remains an 
insulator from $t=-\infty$ to $t=0$. 
When we take the static limit ($\omega\rightarrow 0$) 
, $\bar{J}_{{\rm el},\mu}$ given in eq.(\ref{2-0-3}) is in 
fact identical to 
$\delta\vec{P}_{\rm{el}}$ given in eq.(\ref{2-4}).

\section{fictitious magnetic charge}
In this Appendix, we will prove that the 
doubly degenerate point defined in eq.(\ref{2-12}) 
is identical to the fictitious magnetic charge 
associated with the flux defined 
in eqs.(\ref{2-8}) and (\ref{2-9}).
%
Namely, we will show that the surface-integral of the 
flux over an closed surface enclosing this 
doubly degenerate point becomes quantized to be $\pm2\pi$.

First we will consider the surface-integral of the flux 
over an infinitesimally small box ${\it{v}}
\equiv\{(K_{x},K_{y},K_{z})|(K_{x},K_{y},K_{z})
\in[-\delta_{x},\delta_{x}]\times
[-\delta_{y},\delta_{y}]\times
[-\delta_{z},\delta_{z}]\}$ enclosing 
this degenerate point, i.e.  
\begin{eqnarray}
\int\int_{\partial {\it{v}}}d\vec{S}\cdot\vec{B}_{n}(\vec{K}).\label{11-1}
\end{eqnarray}
 Here we take the vector $d\vec{S}$ to be directed   
from inside to outside of this box. 
When we define a new coordinate: 
\begin{eqnarray}
\left[\begin{array}{c}
K^{\prime}_{x} \\
K^{\prime}_{y} \\
K^{\prime}_{z} 
\end{array}\right]
&=&\left[\begin{array}{ccc}
	1&& \\
	&{\rm{sign}}({\rm{det}}V)& \\
	&&1
	\end{array}\right]\cdot[V]\cdot
\left[\begin{array}{c}
K_{x} \\
K_{y} \\
K_{z} 
\end{array}\right], \nonumber \\
&\equiv&[T]\cdot\left[\begin{array}{c}
K_{x} \\
K_{y} \\
K_{z} 
\end{array}\right], \nonumber 
\end{eqnarray}
the 2$\times$2 reduced Hamiltonian given in eq.(\ref{2-12}) 
can be rewritten as follows, 
\begin{eqnarray} 
\mbox{\boldmath{$[H]$}}_{2\times2}=
K^{\prime}_{x}[{\mbox{\boldmath{$\sigma$}}}_{x}]+
{\rm{sign}}({\rm{det}}V)\cdot K^{\prime}_{y}
[{\mbox{\boldmath{$\sigma$}}}_{y}]+
K^{\prime}_{z}[{\mbox{\boldmath{$\sigma$}}}_{z}], \label{11-2}
\end{eqnarray}
where we omitted those terms which are proportional to a unit matrix.
Then, accompanied with this coordinate transformation, 
a following flux can be introduced,
\begin{eqnarray}
\vec{B}^{\prime}_{n}(\vec{K}^{\prime})
=-i\nabla_{\vec{K}^{\prime}}\times\langle n,\vec{K}^{\prime}|
\vec{\nabla}_{\vec{K}^{\prime}}
|n,\vec{K}^{\prime}\rangle,\label{11-2-1}
\end{eqnarray}
whose relation to the flux in the old coordinate, i.e. 
$\vec{B}_{n}(\vec{K})$, are given as follows:
\begin{eqnarray}
B_{n,\mu}(\vec{K}) &=& 
{\rm{det}}T\sum_{\nu}[T^{-1}]_{\mu,\nu}
B^{\prime}_{n,\nu}(\vec{K}^{\prime}),\nonumber \\
&=&\sum_{\nu}\tilde{a}_{\nu,\mu}B^{\prime}_{n,\nu}(\vec{K}^{\prime}).
\nonumber
\end{eqnarray}  
Here $\tilde{a}_{\nu,\mu}$ denotes the ($\nu$,$\mu$)-th cofactor 
of the $3\times3$ matrix $[T]$. 

We will show in the following that 
eq.(\ref{11-1}) can be rewritten into the r.h.s. 
of eq.(\ref{11-4}) below. The integral in eq.(\ref{11-1})
 consists of the surface-integral over the 6 rectangulars.
When we focus on one of them,  e.g. the rectangular 
$X_{+}\equiv\{(K_{x},K_{y},K_{z})
|K_{x}=\delta_{x},
(K_{y},K_{z})\in[-\delta_{y},\delta_{y}]\times
[-\delta_{z},\delta_{z}]\}$, the integral over this rectangular
 can be transformed as follows;
\begin{eqnarray}
&&\int\int_{X_{+}}dK_{y}dK_{z}B_{n,x}\nonumber \\
&&=\int\int_{T(X_{+})}d\vec{S}^{\prime}\cdot
\vec{B}^{\prime}_{n}(\vec{K}^{\prime}).\label{11-3}
\end{eqnarray}
Here the direction of $d\vec{S}^{\prime}$ was also taken 
so that it penetrates the parallelogram $T(X_{+})$ from  
inside to outside of  
the box $T({\it{v}})$. This relation results from   
the following mathematics; 
\begin{eqnarray}
&&dK_{y}dK_{z}B_{n,x}\nonumber \\
&&=dK_{y}dK_{z}(\tilde{a}_{xx}B^{\prime}_{n,x}
+\tilde{a}_{yx}B^{\prime}_{n,y}+\tilde{a}_{zx}B^{\prime}_{n,z}),\nonumber \\
&&=(\left[\begin{array}{c}
	T_{xy} \\
	T_{yy} \\
	T_{zy} 
	\end{array}
\right]dK_{y} \times
\left[\begin{array}{c}
	T_{xz} \\
	T_{yz} \\
	T_{zz} 
	\end{array}
\right]dK_{z})\cdot
\left[\begin{array}{c}
	B^{\prime}_{n,x} \\
	B^{\prime}_{n,y} \\
	B^{\prime}_{n,z} 
	\end{array}
\right],\nonumber \\
&&=d\vec{S}^{\prime}\cdot\vec{B}^{\prime}_{n},\nonumber  
\end{eqnarray}
where we used the fact that the parallelogram 
$T(X_{+})$  in eq.(\ref{11-3}) is spanned 
by $\left[ T_{xy},T_{yy},T_{zy}\right]^{t}dK_{y}$ 
and $\left[T_{xz},T_{yz},T_{zz}\right]^{t}dK_{z}$ 
and that the 3$\times$3 matrix $[T]$ is positive definite 
by definition. The other surface-integrals over remaining 5 faces 
can be also transformed  in a similar way. As a result, 
in the $K^{\prime}$-coordinate, eq.(\ref{11-1}) reads:
\begin{eqnarray}
\mbox{eq.(\ref{11-1})}&=&
\int\int_{\partial T({\it{v}})}d\vec{S}^{\prime}\cdot
\vec{B}^{\prime}_{n}(\vec{K}^{\prime}),\nonumber \\
&=&\int\int_{S}
d\vec{S}^{\prime}\cdot\vec{B}^{\prime}_{n}(\vec{K}^{\prime}), 
\label{11-4}
\end{eqnarray}
where $S$ is the sphere 
enclosing the origin, i.e. $\vec{K}^{\prime}=0$.  

By using the spherical coordinate; 
$(K^{\prime}_{x},K^{\prime}_{y},K^{\prime}_{z})
=K^{\prime}(\sin\theta\cos\phi
,\sin\theta\sin\phi,\cos\theta)$, 
the eigenvector of eq.(\ref{11-2})
 can be given as follows,  
\begin{eqnarray}
|n,\vec{K}^{\prime}\rangle=\left\{\begin{array}{rl}
\left[\begin{array}{c}
	\cos\frac{\theta}{2} \\
	\sin\frac{\theta}{2}e^{i\phi\cdot{\rm{sign}}({\rm{det}}V)}
	\end{array}\right],& \quad \mbox{for $E_{n}=K^{\prime}$} \\
\left[\begin{array}{c}
	\sin\frac{\theta}{2}e^{-i\phi\cdot{\rm{sign}}({\rm{det}}V)} \\
	-\cos\frac{\theta}{2}
	\end{array}\right].& \quad \mbox{for $E_{n}=-K^{\prime}$}
\end{array}\right.  \nonumber    
\end{eqnarray}  
Accordingly, the flux defined in eq.(\ref{11-2-1}) 
can be calculated easily, 
\begin{eqnarray}  
\vec{B}^{\prime}_{n}(\vec{K}^{\prime})
&=& \frac{1}{{K^{\prime}}^{2}\sin\theta}(
\frac{\partial}{\partial \theta}\langle n, \vec{K}^{\prime}|
\frac{\partial}{\partial \phi}| n, \vec{K}^{\prime}\rangle
-{\rm{c.c.}})\frac{\vec{K}^{\prime}}{K^{\prime}},\nonumber \\  
&=& \pm\frac{1}{2{K^{\prime}}^{2}}
\cdot{\rm{sign}}({\rm{det}}V),\label{11-5} 
\end{eqnarray}   
where the sign $+$ is for the case where the $n$-th band is the upper band 
and the $-$ sign is for the lower band. 
Substituting eq.(\ref{11-5}) into the r.h.s. of 
eq.(\ref{11-4}), we get  
\begin{eqnarray}  
&&\int\int_{\partial {\it{v}}}d\vec{S}\cdot\vec{B}_{n}(\vec{K})\nonumber \\
&&=\pm 2\pi\cdot{\rm{sign}}({\rm{det}}V), \label{11-6} 
\end{eqnarray}
where $\pm$ correspond to those in eq.(\ref{11-5}). 
To summarize, in the case of ${\rm{det}}V<0$, 
the flux for the upper band sinks into $\vec{K}=0$, while 
that for the lower band streams from this origin 
with $2\pi$ charge. But in the case of 
${\rm{det}}V<0$, the above relations are reversed.

\section{$\vec{M}_{\rm AF}$  collinear to an arbitrary direction}
In this Appendix, we will argue that a double degeneracy always 
occurs on the line; $\vec{M}_{\perp}=k_{y}=0$
in the $k_{x}-k_{y}-\vec{M}_{\perp}$ space. As we 
argued in eq.(\ref{3-11}), the following two Bloch wave
functions are energetically degenerated:
\begin{eqnarray}
&&\langle a,\mu,\alpha|n(\vec{M}_{\perp}=0,\vec{k})\rangle =
\nonumber \\
&&\sum_{b,\beta}\left[\mbox{\boldmath{$\sigma$}}_{x}
\right]_{ab}
\left[-i\mbox{\boldmath{$\sigma$}}_{y}
\right]_{\alpha\beta}
\langle b,\mu,\beta|n(\vec{M}_{\perp}=0,-\vec{k})
\rangle^{\ast}\label{12-1}, 
\end{eqnarray}
since our 2D system with the collinear AF mean field is 
invariant under the time reversal operation combined 
with the the spatial translation, i.e. $\{R|{\rm a}_{y}\}$.

In addition to this, the periodic part of the 
Bloch function at $(k_x,k_y)$ and 
that of $(-k_x,k_y)$ are identical to each other, 
which are also energetically degenerated:
\begin{eqnarray} 
\langle a,\mu,\alpha|n(\vec{M}_{\perp},k_x,k_y) \rangle 
=\langle a,\mu,\alpha|n(\vec{M}_{\perp},-k_x,k_y) 
\rangle. \label{12-2} 
\end{eqnarray}
We can easily verify eq.(\ref{12-2}), when we 
write down our Hamiltonian in the momentum representation:  
$\sum [H(\vec{k},\vec{M}_{\perp})]_
{(a,\mu,\alpha|b,\nu,\beta)}
f^{\dagger}_{\vec{k},a,\mu,\alpha}f_{\vec{k},b,\nu,\beta}$.   
Namely, $[H(\vec{k},\vec{M}_{\perp})]$ is composed from 
the following five parts:
\begin{eqnarray}
[H]&=&[H_{\rm CF}] + [H_{{\rm K},y}] +  
[H_{\rm LS}] + [H_{\rm MF}] +  [H_{{\rm K},x}], 
\nonumber \\
\left[H_{\rm CF}\right]
&=& \epsilon_{\mu}\delta_{ab}\delta_{\mu\nu}
\delta_{\alpha\beta}, \nonumber \\
\left[H_{{\rm K},y}\right]
&=& 2\cos(k_{y}{\rm{a}}_{y})
\left[\mbox{\boldmath{$\sigma$}}_{x}
\right]_{ab}[\mbox{\boldmath{$t$}}^{y,s}]
_{\mu\nu}\delta_{\alpha\beta}
\nonumber \\
&&\hspace{0.3cm} +\ \ 2\sin(k_{y}{\rm{a}}_{y})
\left[\mbox{\boldmath{$\sigma$}}_{y}
\right]_{ab}
[\mbox{\boldmath{$t$}}^{y,a}]
_{\mu\nu}
\delta_{\alpha\beta},\nonumber \\
\left[H_{\rm LS}\right]
&=&\sum_{\lambda=x,y,z}
\delta_{ab}
[\mbox{\boldmath{$L$}}_{\lambda}]_{\mu\nu}
[\mbox{\boldmath{$\sigma$}}_{\lambda}]_{\alpha\beta}, 
\nonumber \\
\left[H_{\rm MF}\right]
&=&\sum_{\lambda}\Bigl\{(\vec{M}_{\rm AF})_{\lambda}
\left[\mbox{\boldmath{$\sigma$}}_{z}\right]_{ab} 
+ (\vec{M}_{\perp})_{\lambda}
\delta_{ab}\Bigr\}\delta_{\mu\nu}
[\mbox{\boldmath{$\sigma$}}_{\lambda}]_{\alpha\beta}, 
\nonumber \\
\left[H_{{\rm K},x}\right]
&=&
2{\bf{\Delta}}\cos(k_{x}{\rm{a}}_{x})
\left[\mbox{\boldmath{$\sigma$}}_{x}\right]_{ab}
[\mbox{\boldmath{$t$}}^{x,s}]_{\mu\nu}
\delta_{\alpha\beta},\nonumber 
\end{eqnarray}
where $a,b$ denote the sublattice indice and $\mu,\nu$  
and $\alpha,\beta$ are those for the orbital and 
spin respectively. 
Since we didn't take into account the {\it antisymmetric} 
transfer integrals along the $x$-direction, 
$[H(\vec{k},\vec{M}_{\perp})]$ is clearly  
invariant under the sign change of $k_{x}$.  
Therefore its eigenvector at $(k_{x},k_{y})$ is identical 
to that of  $(-k_{x},k_{y})$. 

Eqs.(\ref{12-1},\ref{12-2}) indicate that the following 
two Bloch functions are energetically degenerated along  
$k_{y}=0$ when the ferromagnetic moment vanishes 
($\vec{M}_{\perp}=0$);
\begin{eqnarray}
&&\langle a,\mu,\alpha|n(\vec{M}_{\perp}=0,k_x,k_y)\rangle =
\nonumber \\
&&\sum_{b,\beta}
\left[\mbox{\boldmath{$\sigma$}}_{x}\right]_{ab}
\left[-i\mbox{\boldmath{$\sigma$}}_{y}\right]_{\alpha\beta}
\langle b,\mu,\beta|n(\vec{M}_{\perp}=0,k_x,-k_y)
\rangle^{\ast}\label{12-3}. 
\end{eqnarray}

\section{$\vec{M}_{\rm AF}$  collinear to the $z$-direction}
In this Appendix, we will argue that when 
$\vec{M}_{\rm AF}$ is collinear to the $z$-direction, 
another double degeneracy occurs at X point, i.e.  
$(k_{x},k_{y})=(\frac{\pi}{2{\rm{a}}_{x}},\pm\frac{\pi}{2{\rm{a}}_{y}})
\equiv\vec{k}_{X}$, irrespective of $\vec{M}_{\perp}$. Namely, 
a doubly degenerate surface appears in the 4D parameter space 
space spanned by $k_x$,$k_y$ and $\vec{M}_{\perp}$. 

When $\vec{M}_{\perp}=0$ with $\vec{M}_{\rm AF}$ 
collinear to the $z$-direction, the system is 
invariant under the following symmetry operations;  
\begin{eqnarray}
E,\{\sigma_{y}|{\rm{a}}_{y}\},IC_{2z}
,\{C_{2x}|{\rm{a}}_{y}\}, \nonumber 
\end{eqnarray}
in addition to $\{R|{\rm{a}}_{y}\}$. Here
$I$,$\sigma_{y}$ and $C_{2x(z)}$ represent
the spatial inversion, mirror with respect to
the $y=0$ plane and  the spatial rotation by $\pi$
around the $x(z)$-axis respectively. Then the $k$-group 
at X point is composed of a unitary subgroup 
$G_{X}\equiv T_{\vec{R}_{l}}\times\{E,IC_{2z}\}$
and an anti-unitary subgroup
$\{R|{\rm{a}}_{y}\}\times T_{\vec{R}_{l}}\times\{E,IC_{2z}\}$.
$T_{\vec{R}_{l}}$ denotes the spatial translation by 
$\vec{R}_{l}\equiv n({\rm a}_x +{\rm a}_y) + 
m({\rm a}_x -{\rm a}_y)$, where $n$ and $m$ take 
an arbitrary integer. Then the irreducible 
representation of the unitary subgroup
$T_{\vec{R}_{l}}\times\{E,IC_{2z}\}$ becomes 
$e^{i\vec{k}_{X}\cdot\vec{R}_{l}}\Gamma_{C_{1h}}(\{E,IC_{2z}\})$,
where $\Gamma_{C_{1h}}$ denotes the representation of the
monoclinic point group $C_{1h}$.  The character table for the  
monoclinic point group is given in Table.I,  
\begin{table}[htbp]
\begin{center}
\begin{tabular}{|c||cccc|}
\hline
$C_{1h}$&$E$&$\bar{E}$&$IC_{2z}$&$I\bar{C}_{2z}$ \\ \hline
$\Gamma_{3}$&$1$&$-1$&$-i$&$i$ \\ 
$\Gamma_{4}$&$1$&$-1$&$i$&$-i$ \\ \hline
\end{tabular}
\caption{Character table for double-valued representation of the 
monoclinic point group $C_{1h}$.  
$I\bar{C}_{2z}\equiv\bar{E}IC_{2z}$} 
\end{center}
\end{table}
where we only take the double-valued representation since we
treat the spinful Hamiltonian with spin-orbit interations.
$\bar{E}$ in Table.I represents the $2\pi$ rotation in
the spin-$\frac{1}{2}$ space whose character is always $-1$. 
The dimensionality of these two representation are doubled due 
to the presence of the anti-unitary subgroup, 
$\{R|{\rm{a}}_{y}\}\times T_{\vec{R}_{l}}\times\{E,IC_{2z}\}$.    
Namely, since the Wigner's criterion\cite{Inui} reads  
\begin{eqnarray}
&&\sum_{u\in G_{X}, \bar{E}G_{X}}
\chi^{X}
\Bigl((\{R|\vec{\rm{a}}_{y}\}\cdot u)^{2}\Bigr)\nonumber \\
&&=2\sum_{\vec{R}_{l}}
e^{2i\vec{k}_{X}\cdot(\vec{\rm{a}}_{y}+\vec{R}_{l})}
\Bigl(\Gamma_{C_{1h}}(R^{2})+\Gamma_{C_{1h}}(R^{2}C^{2}_{2z})
\Bigr),\nonumber \\
&&=2\sum_{\vec{R}_{l}}
e^{2i\vec{k}_{X}\cdot(\vec{\rm{a}}_{y}+\vec{R}_{l})}
(\Gamma_{C_{1h}}(\bar{E})+\Gamma_{C_{1h}}(E))=0, \label{13-1}
\end{eqnarray}
only 2-dimensional representations are allowed at $X$ point 
irrespective of whether $\Gamma_{C_{1h}}$ is chosen to be 
$\Gamma_3$ or $\Gamma_4$. Here $\chi^{X}$ in eq.(\ref{13-1}) 
denotes the character of 
the irreducible representation of the $k$-group at $X$-point. 

These degeneracies still remain even when we violate the time-reversal
symmetry $\{R|{\rm{a}}_{y}\}$ by introducing finite $\vec{M}_{\perp}$.
In the case of $\vec{M}_{\perp}\ne 0$, 
the system is still invariant under the 
$T_{\vec{R}_{l}}\times\{E,\{RIC_{2z}|{{\rm{a}}}_{y}\}\}$.
Then the degeneracy at $X$ point becomes doubled due to  
its anti-unitary element $\{RIC_{2z}|{{\rm{a}}}_{y}\}$:
\begin{eqnarray}
&&\sum_{u=T_{\vec{R}_{l}}\times\{E,\bar{E}\}}
\chi^{X}((\{RIC_{2z}|{{\rm{a}}}_{y}\}\cdot u)^{2})\nonumber \\
&&=2\sum_{\vec{R}_{l}}e^{2i\vec{k}_{X}\cdot(\vec{R}_{l}+\vec{\rm{a}}_{y})}
=-2N, \nonumber 
\end{eqnarray}
where $2N$ is the total number of the elements of 
the unitary k-group $T_{\vec{R}_{l}}\times\{E,\bar{E}\}$.
To summarize, we always have a double degeneracy at X point 
regardless of $\vec{M}_{\perp}$.


\begin{thebibliography}{9}
\bibitem{symmetry}The time reversal operator acts on our Hamiltonian, 
$\hat{H}=H_{\left(i,\mu,\alpha|j,\nu,\beta\right)}
f^{\dagger}_{i,\mu,\alpha}f_{j,\nu,\beta}$ in the follow way;
\begin{eqnarray}
H_{\left(i,\mu,\alpha|j,\nu,\beta\right)}
&\rightarrow&\sum_{\mu^{\prime}\mu^{\prime}}
\sum_{\alpha^{\prime}\beta^{\prime}}
[R^{(t)}]^{-1}
_{\mu\mu^{\prime}}
[\sigma_{y}]_{\alpha\alpha^{\prime}}\nonumber \\
&&H_{\left(i,\mu^{\prime},\alpha^{\prime}
|j,\nu^{\prime},\beta^{\prime}\right)}^{*}
[R^{(t)}]_{\nu^{\prime}\nu}
[-i\sigma_{y}]_{\beta^{\prime}\beta}, \label{15-1}
\end{eqnarray}
while the spatial rotation around $\vec{n}$ by $\theta$ ,i.e., 
$C(\vec{n},\theta)$ acts as follows, 
\begin{eqnarray}
H_{\left(i,\mu,\alpha|j,\nu,\beta\right)}&\rightarrow&
\sum_{\mu^{\prime},\mu^{\prime}}
\sum_{\alpha^{\prime},\beta^{\prime}}[e^{-i\theta\vec{n}
\cdot\vec{L}}]_{\mu\mu^{\prime}}
[e^{-i\theta\vec{n}\cdot\vec{S}}]
_{\alpha\alpha^{\prime}}\nonumber  \\
&&H_{\left(C(\vec{n},\theta)^{-1}(i),\mu^{\prime},\alpha^{\prime}
|C(\vec{n},\theta)^{-1}(j),\nu^{\prime},\beta^{\prime}\right)}
\nonumber \\
&&[e^{i\theta\vec{n}\cdot\vec{L}}]_{\nu^{\prime}\nu}
[e^{i\theta\vec{n}\cdot\vec{S}}]
_{\beta^{\prime}\beta}. \label{15-2}
\end{eqnarray}
Here the indice $i,j$ are the site indice (not only magnetic 
unit cell indice but also sublattice indice are included), 
$\mu$ and $\nu$ are orbital indice and $\alpha,\beta$ are those for spin.
As for the operators on the orbital space, 
when the basis functions for d-orbital 
indice $\mu$ and $\nu$ were taken as    
$|L^2=2(2+1),L_{z}=2\rangle,|6,1\rangle,|6,0\rangle
(\equiv|6,0\rangle^{\ast}),|6,-1\rangle(\equiv|6,1\rangle^{\ast}),
|6,-2\rangle(\equiv|6,2\rangle^{\ast})$, we replaced   
$[R^{(t)}]$ in Eq.(\ref{15-1}) by   
\begin{eqnarray}
\left[\begin{array}{ccccc}
0&0&0&0&1 \\
0&0&0&1&0 \\
0&0&1&0&0 \\
0&1&0&0&0 \\
1&0&0&0&0
\end{array}\right]\nonumber .
\end{eqnarray}
However, in the convention of this paper, we take 
the basis functions for d-orbital as real functions, which means that    
$[R^{(t)}]$ must be identity.
Accordingly, $5\times5$ matrices $[\vec{L}]$ 
in the spatial rotational operators should be taken as in eq.(\ref{3-9}).
As for the operators on the spin space, 
we take $\vec{S}
=\frac{1}{2}\vec{\sigma}$.

$C(\vec{n},\theta)^{-1}(i)$ in Eq.(\ref{15-2}) represents the site 
which transports onto the site $i$ when we apply 
the spatial rotation $C(\vec{n},\theta)$. 
In contrast to the above operations, the spatial  
inversion $I$ affects only on the site index $i$ because the $d$-orbitals 
are always even under $I$,  
\begin{eqnarray}
H_{(i,\mu,\alpha|j,\nu,\beta)}&\rightarrow&
H_{(I(i),\mu,\alpha|I(j),\nu,\beta)}.\nonumber
\end{eqnarray}
\bibitem{anote1} From the symmetry considerations, 
Dzyaloshinskii-Moriya (DM) interaction turns out to have 
no staggered components, while the uniform component 
of the DM vector is allowed to head for 
the z-direction. However, as far as canted AF spin 
configurations are concerned, uniform components of 
DM vector do not matter.
\bibitem{anote2} Magnitudes of symmetric parts of 
anisotropic exchange interactions $\Delta J$ are 
usually proportional to square of the 
spin-orbit coupling constant and thus very small.
\bibitem{note6} Effects of depolarization (depolarizing) 
 fields on the macroscopic electronic polarization (MEP) 
have been phenomenologically argued in the review article 
listed above\cite{resta}.
\bibitem{note1}
$\bar{J}_{\rm el}(x)$ is defined by the time-dependent 
ground state wavefunction $|\Psi(t) \rangle $:  
\begin{eqnarray}
\bar{J}_{\rm el}(x) \equiv \int_{-\infty}^{0} dt   
\langle \Psi(t) |J_{\rm el}(x) | \Psi(t) \rangle. 
\end{eqnarray} 
This ground state wave function $|\Psi(t) \rangle $ 
is now adiabatically evolved according to the time-dependent Hamiltonian 
given in eq.(\ref{2-0-01}), where an infinitesimal 
perturbation $\delta \vec{\varphi}$
is introduced during the interval $t\in [-\infty,0]$.
\bibitem{note2} The inhomogeneity of the current distribution around 
sample boundaries, i.e. $x+\pm\frac{L}{2}$, does not matter 
with our arguments, when $L$ is taken to be sufficiently large.
\bibitem{note2-1} In order to take into account the spatial variation 
of this depolarization field accurately, we need to slove  
the following self-consistent equation  for $\bar{J}_{\rm el}(x)$ 
instead of eq. (\ref{2-16}):
\begin{eqnarray}
\bar{J}_{\rm el}(x) = -\frac{i}{V}\sum_{m\ne 0}
\biggl[\frac{\langle \Psi_{0}|J_{\rm el}(x)|\Psi_{m} \rangle 
\langle \Psi_{0}|\delta H_{\rm M.F.}|\Psi_{m}\rangle}
{(E_{m}-E_{0})^{2}} -{\rm c.c.}\biggr]  \nonumber \\
+ \frac{ie}{\epsilon_{0}V}\sum_{j=1}^{N} \sum_{m\ne 0}
\biggl[\frac{\langle \Psi_{0}|J_{\rm el}(x)|\Psi_{m} \rangle 
\langle \Psi_{0}|x_{j}|\Psi_{m}\rangle}
{(E_{m}-E_{0})^{2}} -{\rm c.c.}\biggr] \bar{J}_{\rm el}(x_{j}). 
\end{eqnarray}
Then, by integrating $\bar{J}_{\rm el}(x)$ w.r.t. $x$, we  
obtain $\delta P_{\rm el}$ which is exact at the mean-field level.
\end{thebibliography}
\end{document}